\documentclass{ws-ijmpd}
\newcommand{\aaa}{{\it Astron. Astrophys.}}
\newcommand{\apj}{{\it Astrophys. J.}}
\newcommand{\araa}{{\it Ann. Rev. Astron. Astrophys.}}
\newcommand{\JCAP}{{\it J. Cos. Astropart. Phys.}}
\newcommand{\MNRAS}{{\it Mon. Not. R. Astron. Soc.}}
\newcommand{\Nature}{{\it Nature.}}
\newcommand{\NPB}{{\it Nul. Phys. B.}}
\newcommand{\PLB}{{\it Phys. Lett. B.}}
\newcommand{\PRL}{{\it Phys. Rev. Lett.}}
\newcommand{\PRD}{{\it Phys. Rev. D.}}
\newcommand{\RMP}{{\it Rev. Mod. Phys.}}
\begin{document}

\markboth{Y. Z. Fan , B. Zhang \& J. Chang}
{$e^\pm$ Excess in the Cosmic Ray Spectrum and Possible Interpretations}

\catchline{}{}{}

\title{$e^\pm$ Excesses in the Cosmic Ray Spectrum and Possible Interpretations
}

\author{\footnotesize Yi-Zhong Fan
}

\address{Department of Physics and Astronomy, University of Nevada, Las Vegas, NV 89119, USA;\\
Purple Mountain Observatory, Chinese Academy of Sciences, 210008, Nanjing, China
}

\author{Bing Zhang}
\address{Department of Physics and Astronomy, University of Nevada, Las Vegas, NV 89119, USA
}

\author{Jin Chang}
\address{Purple Mountain Observatory, Chinese Academy of Sciences, 210008, Nanjing, China
}

\maketitle


\begin{history}
\received{Day Month Year}
\revised{Day Month Year}
\comby{Managing Editor}
\end{history}

\begin{abstract}

The data collected by ATIC, PPB-BETS, FERMI-LAT and HESS
all indicate that there is an electron/positron excess in the cosmic
ray energy spectrum above $\sim$ 100 GeV, although different instrumental
teams do not agree on the detailed spectral shape.
PAMELA also reported a clear excess feature of the positron fraction
above several GeV, but no excess in anti-protons.
Here we review the observational status and theoretical models of this
interesting observational feature. We pay special attention to various
physical interpretations proposed in the literature,
including modified supernova remnant models for the $e^\pm$
background, new astrophysical sources, and new physics (the dark
matter models).
We suggest that although most models can make a case to interpret the
data, with the current observational constraints the dark
matter interpretations, especially those invoking annihilation,
require much more exotic assumptions than some astrophysical
interpretations.
Future observations may present some ``smoking-gun'' observational
tests to differentiate among different models and to identify
the correct interpretation to the phenomenon.

\end{abstract}

\section{Introduction}
Cosmic rays (CRs) are energetic particles originating from outer space
that impinge on Earth's atmosphere. Almost $90\%$ of all the incoming
cosmic ray particles are protons, about $10\%$ are helium nuclei
(alpha particles), and slightly under $1\%$ are heavier elements and
electrons\cite{Ginzburg96PhysU}. The history of the CR studies
may be rooted in 1901 when two groups came to the same conclusion that
pure air in a closed vessel possessed some electrical
conductivity\cite{Elster1901,Wilson1901}, while no visible sources of
air ionization were observed.  At that time it was already known that
X-rays and radioactivity were factors contributing to the enhanced
electrical conductivity of gases. Therefore, the observed effect of
``dark current" caused by residual ionization of the air, was then
regarded as being associated with radioactive contamination both in the
air and in the environment\cite{Ginzburg96PhysU}.  Wilson speculated
that the residual ionization was due to certain highly penetrating
radiation coming from outside the Earth's atmosphere\cite{Wilson1901b}.
Such a speculation was convincingly confirmed
by Hess\cite{Hess1912} and Kolhorster\cite{Kolhorster1913}
about 10 years later. August 7, 1912, the date of Hess' most
successful balloon flight among the total of ten, is generally regarded
as the date of the discovery of cosmic rays.  The discovery of cosmic
rays opens a new window not only for high energy astrophysics but also
for high energy physics. This is because prior to the constructions of
modern particle accelerators, high energy particles could only be
observed in cosmic rays. Between 1932 and 1953, many new particles,
including positrons, $\mu^{\pm}$-leptons,
$\pi^{\pm}$-mesons, $K^{+}-$ and $K^{0}-$mesons, $\Lambda^{0}-$,
$\Xi^{-}-$ and $\Sigma^{+}-$hyperons, were all discovered in use of
cosmic rays (Ref.\refcite{Ginzburg96PhysU} and the references therein).

A thorough review of CRs can be found in many excellent articles
and books\cite{Ginzburg64book,Meyer69ARAA,Hillas84ARAA,Nagano00RMP},
and is beyond the scope of this review.
Here we will focus on a small fraction
of these high energy particles, the electron/positron ($e^{\pm}$) CRs.
Although CRs had been discovered in 1912, CR electrons reaching the
earth at energies above a few hundred MeV had not been convincingly
identified until 1961\cite{Earl61PRL,Meyer61PRL}.
The first positron-to-electron ratio in CRs up to an energy $\sim 1$
GeV was later reported in 1964\cite{De-Shong64PRL}. Most CR electrons
are likely from the supernova remnants while the CR positrons are
mainly produced through hadronic processes when CR protons collide
with intergalactic hydrogen.
In such a scenario there should be no prominent feature in the
TeV energies in the electron/positron total spectrum, and the
positron-to-electron ratio ${\cal R}\equiv
\Phi_{e^{+}}/(\Phi_{e^{+}}+\Phi_{e^{-}})$ should drop with energy
monotonously\cite{Strong07Rev}.  A rise of the positron-to-electron
ratio above $\sim 10$ GeV was observed in
1974\cite{Buffingston74PRL}\footnote{In 1965,
a group reported ${\cal R}\sim 0.8$ at
$10-30$ GeV, implying an excess of positron CRs\cite{Daniel65PRL}. It
was however only based on a total of 13 events and therefore highly
uncertain \cite{Daniel70SSRev}. If one plots the single data of
Ref.\refcite{Agrinier69} (at $\sim 10$ GeV) published in 1969 together
with some other data at lower energies reported before, one would also
get a positron excess\cite{Muller87}.}.  Such a tendency was
confirmed by several balloon flights, including the 1976 flight by
the New Mexico State University group launched from
Palestine, Texa in 1976\cite{Golden87AA}, and 1984 flight by the
University of Chicago group launched from Hawaii\cite{Muller87},
and in particular, by the recent PAMELA
mission\cite{pamela-e,pamela-e1} (see section \ref{sec:Observation}
for extensive discussion). In 1987 it was already clear that the
rising behavior of ${\cal R}$ above $\sim 10$ GeV is not predicted by
calculations of CR propagation\cite{Muller87,Protheroe82} and may
imply an excess of CR positrons.  Recently, ATIC\cite{atic},
PPB-BETS\cite{ppb-bets} and Fermi-LAT\cite{fermi} reported
electron/positron total spectrum up to TeV and found a hardening/bump
in the energy range from 100 GeV to 1 TeV. In the conventional
approach, the injection spectrum of the electrons (positrons) is
taken as a single power-law. Since diffusion and electron/positron
cooling are more efficient in higher energies, one would expect that
the spectrum should soften with energy. Therefore the unexpected
spectrum hardening observed by these detectors strongly suggest
another excess (besides the positron fraction excess).
Interestingly this total $e^\pm$ excess is consistent with
extrapolating the positron fraction excess to high energies.
These interesting features draw a lot of attention, and various
physical origins have been explored. In the literature, new
astrophysical sources or even new physics (dark matter) are
widely suggested. Some more conservative authors argue that
the excesses are simply due to inadequate accounts on
the electron/positron CR background 
in previous modeling. The situation is unclear. Several reviews
appeared recently but are mainly focused on dark matter models
\cite{He09IJPA,Boezio09NJP}. In this review we plan to present
a more complete overview of this interesting observational feature
and various physical interpretations.

The structure of this review is as follows. We first discuss the
observational aspects of the $e^\pm$ excesses in Section 2. We then
review the cosmic ray propagation models and the general cosmic
sources of $e^\pm$ in Section 3. In the next three sections (4-6),
we discuss three proposed physical origins of the observed $e^\pm$
excess, including the modified supernova remnant (SNR) models of
$e^\pm$ background, new astrophysical sources, and new physics.
We summarize the strengths/weaknesses of various models in Section
7 with a commentary on the prospects.

\section{Observations}\label{sec:Observation}
Different from the CR protons/nuclei, the TeV electrons/positrons
suffer significant energy loss and cannot reach us if the source is
not nearby (i.e., $\leq 1$ kpc). Accurate measurements of the high-energy
electron/positron CRs then provide a unique opportunity to probe the
origin and propagation of CRs in the local interstellar medium and to
constrain models of the diffuse gamma-ray emission. Observations of
electrons have been notoriously difficult because of their low
intensity requiring, in particular at high energies, rather large
detectors, and because of the need of effective discrimination against
proton-induced background. So far the high-energy electron spectrum
was measured by some balloon-borne experiments (in addition to those
already mentioned in the introduction,
some others can be found in Refs.
\refcite{Critchfield52PRev,Agrinier64PRL,Hartman67ApJ,Fanselow69ApJ,Golden94ApJ,Barbiellini96AA,Golden96ApJ,Barwick97ApJ,Boezio01ApJ,Torii01ApJ}),
some space missions (including IMP-1\cite{Cline64PRL},
IMP-III\cite{Fan68ApJ}, IMP-IV\cite{Simnent69ApJ},
IMP-7\cite{Hurford74ApJ}, AMS-01\cite{ams}, PAMELA\cite{pamela-e} and
Fermi\cite{fermi}), and the ground-based Cherenkov telescope
H.E.S.S.\cite{hess}.

\subsection{Experiments}\label{sec:Experiments}
\subsubsection{Balloon-borne experiments}
We first briefly review the development of technology used
in the experiments.  In two of the very early Balloon-borne
experiments\cite{Critchfield52PRev,Earl61PRL}, the cloud chamber
technology was employed, which was soon abandoned with
the introduction of other more sophisticated detectors, both
visible and electronic
(see Ref.\refcite{Daniel70SSRev} for a review
on the development of technology in the early experiments). In the
two flights at Fort Churchill in 1963\cite{De-Shong64PRL}, the spark
chamber technique was adopted. With the help of one permanent magnet,
electrons and positrons were distinguished. Such an
instrument was later improved\cite{Hartman67ApJ}, through replacing
the liquid Cerenkov detector by a gas Cerenkov detector to enable the
determination of the momentum and charge of electrons up to $5-10$
GeV.  A group from Bombay\cite{Daniel65PRL,Anand68PRL} adopted the
nuclear emulsion stack technique, which measures CR electrons to an
energy above 50 GeV. In order to identify electron showers
efficiently, multiple pulse-height analyses of many counters and time
of flight techniques had been used in 3 high-latitude balloon flights
from Palestine, Texas in 1970 by the Chicago
group\cite{Muller73ApJ}. Later they added a transition radiation
detector, consisting of a sandwich of six radiators and six multiwire
proportional chambers, to improve electron and positron
discrimination\cite{Hartmann77PRL,Tang84ApJ}. In a balloon-borne
magnetic spectrometer experiment\cite{Buffingston74PRL,Buffingston75},
a super-conducting magnetic spectrometer was adopted, and a new
bremsstrahlung-identification technique was developed
to measure the separate $e^{-}$ and $e^{+}$ spectrum in the primary
CRs from 4 to 50 GeV. The emulsion chamber technique was used to
detect CR electrons to an
energy above 1 TeV\cite{Nishimura80ApJ,Kobayashi99ICRC}. In the
experiment of Balloon-borne Electron Telescope with Scintillating
Fibers (BETS), the absolute energy spectrum of electrons was measured
with a highly granulated fiber calorimeter\cite{Torii01ApJ}. Finally
the Advanced Thin Ionization Calorimeter (ATIC) experiments used an
ionization calorimeter to measure the energy deposited by cascades
formed by particles interacting in a thick carbon target \cite{atic}.

Since 1950, dozens of balloon-born experiments dedicated to
electron/positron CR observations have been carried out.
Below we only mention some recent ones that are most relevant to
the main topic of this review, the electron/positron excess.

{\bf Two early charge determining experiments}. The Chicago group
firstly used a permanent magnet to measure the positron-to-electron ratio. In
the energy range $0.2~{\rm GeV}<{\cal E}<10$ GeV,
the ratio ${\cal R}$ was found to decrease with
energy\cite{Fanselow69ApJ}. Adopting a super-conducting
magnetic spectrometer, ${\cal R}$ was measured from 5
GeV to 50 GeV\cite{Buffingston74PRL}. Although scatter is large, an
tendency of increasing ${\cal R}$ for ${\cal E}>10$ GeV is
present. Such a behavior was confirmed by some subsequent groups
(e.g. Ref. \refcite{Muller87,Golden87AA}),
but not by some others.

{\bf The emulation chamber experiments} by a Japanese group
started in 1968. The $e^{+}+e^{-}$ spectrum derived by combining data
from the chambers exposed from 1968 to 1976 was well represented by
$J({\cal E})=1.6\times 10^{-4}({\cal E}/100~{\rm GeV})^{-3.3\pm
0.2}~{\rm m^{-2}~sr^{-1}~s^{-1}~GeV^{-1}}$ in the energy range of
$30-10^{3}$ GeV\cite{Nishimura80ApJ}, which is steeper than the
spectrum $J \propto {\cal E}^{-2.61\pm 0.1}$ obtained in the
early nuclear emulation stack experiment by the Bombay
group\cite{Anand73ICRC}. This confirms some early results,
for example, obtained with an ionization
calorimeter\cite{Earl72JGR,Meegan75ApJ}. The observations in 1996 and
1998 extended the $e^{+}+e^{-}$ spectrum to $\sim 2$
TeV\cite{Kobayashi99ICRC} and might suggest a steepening above
1 TeV.

{\bf High-Energy Antimatter Telescope (HEAT)} is a NASA-supported
program of high-altitude balloon-borne experiments to study antimatter
in the primary cosmic radiation.  The HEAT detector\cite{Barwick97ApJ}
consists of a magnetic spectrometer combined with
a transition radiation detector, an electromagnetic calorimeter, and
time-of-flight scintillators. It can measure the cosmic-ray positron
fraction from 1 to 50 GeV.  A new version of the HEAT instrument,
HEAT-pbar was designed to observe the high-energy cosmic-ray
antiproton flux but it is also suited for the observation of electrons
and positrons at energies below 15 GeV\cite{Beatty04PRL}.  The
combined data from the three HEAT flights indicate a small positron
flux of nonstandard origin above 5 GeV\cite{Beatty04PRL}. The
evidence for the tendency of increasing ${\cal R}$ for ${\cal E}>10$
GeV is however rather weak, consistent with what was found in the Cosmic
AntiParticle Ring Imaging Cherenkov Experiment in 1998\cite{Boezio01ASR}.

{\bf BETS and PPB-BETS.} BETS represents a detector, the Balloon-borne
Electron Telescope with Scintillating fibers, which preserves the
superior qualities of both electronic detectors and emulsion chambers
\cite{Torii01ApJ}. With such a balloon-borne payload, cosmic-ray
electrons were observed in the energy range from 12 to $\sim$100
GeV. The energy spectrum is described by a power-law index of $-3.00
\pm 0.09$, and the absolute differential intensity at 10 GeV is
$0.199\pm 0.015 ~{\rm m^{-2}~ s^{-1}~sr^{-1}~GeV^{-1}}$\cite{Torii01ApJ}.
PPB-BETS is an advanced version of BETS measuring
the electron spectrum up to TeV and was flown by Polar Patrol
Balloon (PPB) in Antarctica.  The results from the PPB-BETS
calorimeter indicates a possible hump in the energy range 100
GeV to 1 TeV in the total spectrum\cite{ppb-bets}, consistent with
that reported by ATIC (see below).

{\bf The Advanced Thin Ionization Calorimeter (ATIC)} is a
balloon-borne instrument flying in the stratosphere over Antarctica to
measure the energy and composition of cosmic rays. ATIC was launched
from the McMurdo Station for the first time in December 2000 and has
since completed three successful flights out of four.  The detector
uses the principle of ionization calorimetry: several layers of the
scintillator bismuth germanate emit light as they are struck by
particles, allowing to calculate the particles' energy. A silicon
matrix is used to determine the particles' electrical
charge\cite{atic}. The main discovery is a spike-like $e^{+}+e^{-}$
spectrum feature in the energy range of $300-800$ GeV\cite{atic}.

\subsubsection{Space station and satellite experiments}

The series of Interplanetary Monitoring Platform (IMP) experiments
measure the low energy electron CRs (${\cal E}<1$ GeV), which are not the
focus of this review. In the following
we only focus on several relevant missions.

{\bf AMS-01} (Alpha Magnetic Spectrometer - 01) is the predecessor to
AMS-02, a detector to be operated on the International Space Station
(ISS) for at least 3 years (http://ams.cern.ch/). The AMS-01 experiment
was flown on the Space Shuttle Discovery for ten days in June, 1998.
It was designed to search for various types of unusual matter by
measuring cosmic rays. The detector consisted of a permanent Nd-Fe-B
magnet, six silicon tracker planes, an anticoincidence scintillator
counter system, the time-of-flight (TOF) system consisting of four
layers of scintillator counters and a threshold aerogel Cerenkov
detector. With these instruments, the high-energy proton, electron,
positron, helium, antiproton and deuterium spectra were accurately
measured. The obtained ${\cal R}$ in the energy range of $1-30$ GeV is
largely consistent with that obtained by HEAT \cite{ams}.

{\bf PAMELA}, the Payload for Antimatter Matter Exploration and
Light-nuclei Astrophysics, is a powerful particle identifier using a
permanent magnet spectrometer with a variety of specialized detectors
(http://pamela.roma2.infn.it/index.php). It is measuring with
unprecedented precision and sensitivity the abundance and energy
spectra of CR electrons, positrons, antiprotons and light
nuclei over a very large range of energy from 50 MeV to hundreds GeV,
depending on the species\cite{pamela-e,pamela-p}. The most exciting
discovery made so far is the unambiguous detection of the increasing
behavior of ${\cal R}$ in the energy range $10-100$ GeV
\cite{pamela-e,pamela-e1}.

{\bf Fermi $\gamma-$ray Space Telescope} is an international and
multi-agency space mission that studies the cosmos in the energy range
10 keV - 300 GeV (http://fermi.gsfc.nasa.gov/). The Large Area
Telescope (LAT) is the main instrument on-board.
Conceived as a
multipurpose observatory to survey the variable gamma-ray sky between
20 MeV and 300 GeV including the largely unexplored energy window
above 10 GeV, it is designed as a low aspect ratio, large
area pair conversion telescope to maximize its field of view and
effective area. The LAT angular, energy and timing resolutions rely on
modern solid state detectors and electronics.  Since electromagnetic
cascades are germane to both electron and photon interactions in
matter, the LAT is also by its nature a detector for electrons and
positrons\cite{fermi}.  The measurement of the CR electron
spectrum from 20 GeV to 1 TeV based on the first half year data
also displays a spectrum hump in the $100-1000$ GeV range,
although the signature is not as prominent as that found by ATIC
and PPB-BETS.

\subsubsection{Ground-based telescopes}
{\bf H.E.S.S}, the High Energy Stereoscopic System, is an array of
four imaging atmospheric Cherenkov telescopes located in the Khomas
Highland of Namibia (http://www.mpi-hd.mpg.de/hfm/HESS/).
It is designed to obtain the Cherenkov
images of cosmic-ray hadrons and electrons, as well as gamma-ray
photons (sorted according to the relative flux). To measure the
spectrum of electrons, the other two components (hereafter background
events) have to be eliminated. First, data taken from the sky regions
known to contain gamma-ray sources are rejected. However, the most
critical aspect of the electron analysis is the efficient rejection of
the hadronic background. A Random Forest approach was used to convert
the image information into a single parameter describing the degree to
which a shower is electron-like, that spans from 0 (for hardons) to 1
(for electrons). By Monte-Carlo simulations, the distribution of this
parameter for simulated hadrons and electrons is determined (See
Figure 1 in Aharonian et al.\cite{hess}). The number of measured
electron showers in each energy band can then be deduced. In this way,
the spectrum is measured from 0.35 - 4 TeV. While the overall electron
flux measured by H.E.S.S. is consistent with the ATIC data within
statistical and systematic errors, the H.E.S.S. data do not confirm a
pronounced peak in the electron spectrum suggested by ATIC and
PPB-BETS\cite{hess1}.

\subsection{Observation results}\label{sec:Results}
\begin{figure}[t]
\begin{center}\includegraphics[width=0.8\columnwidth
]{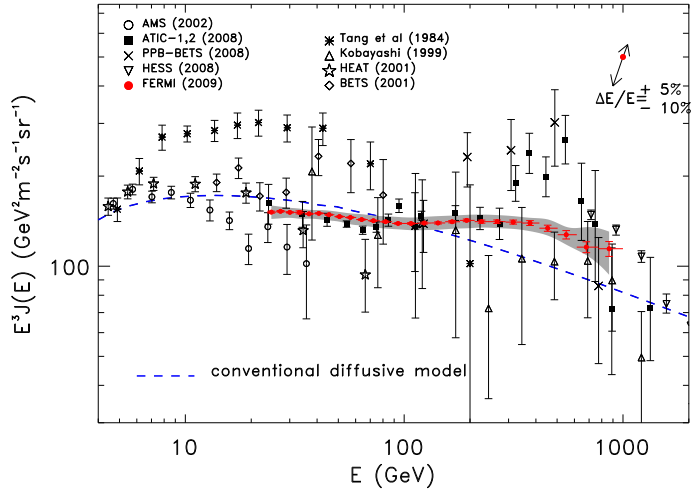}\end{center}
\caption{
\label{data-fermi}
Observational data against the background model estimates for
the $e^+ + e^-$ energy spectrum (from Ref.26).
$J$ is the sum of the fluxes of electrons and positrons.
}
\end{figure}

Prior to 2008, the high energy $e^{+}+e^{-}$ data are consistent
with a featureless power-law spectrum within errors, which is in
agreement with the theoretical predictions from both analytical and
numerical models\cite{Strong07Rev}.  Recently, the ATIC balloon
experiment\cite{atic} discovered a prominent spectral feature at around
600 GeV in the $e^{+}+e^{-}$ spectrum. Instead of a decaying feature
expected by electron/positron cooling, the team discovered a clear
excess of the $e^{+}+e^{-}$ in the energy range of $\sim 300 - 800$
GeV. Such a feature was also
marginally observed by PPB-BETS\cite{ppb-bets}. Furthermore, H.E.S.S.
reported a significant steepening of the electron spectrum above $\sim
1$ TeV\cite{hess,hess1}.  The Fermi-LAT Collaboration later
reported a high-precision measurement of the $e^{+}+e^{-}$
spectrum from 20 GeV to 1 TeV\cite{fermi}. As
shown in Fig.\ref{data-fermi}, their measured $e^{+}+e^{-}$ spectrum
may be fitted by a single power-law
$J=(175.40 \pm 6.09)\varepsilon^{-(3.045\pm
0.008)}~{\rm GeV^{-1}~m^{-2}~s^{-1}~sr^{-1}}$.
Some wiggles are evident if one ignores the systematic
uncertainties. The spectrum reveals a hardening at around 100 GeV and
a steepening above $\sim $ 400 GeV. More specifically, the spectrum
can be fitted by a broken power-law with indices $-3.070 \pm 0.025$
for ${\cal E} < 100$ GeV, $-2.986 \pm 0.031$ for $100 < {\cal E} <
400$ GeV, and $-3.266 \pm 0.116$ for $400 < {\cal E} < 1000$
GeV\cite{Grasso09AP}.
So in general, all detectors reveal {\em an excess} beyond 100 GeV on
the otherwise softening spectrum, although different instrumental
teams do not agree on the detailed spectral shape.

The other independent indication of the presence of a possible
deviation from the standard picture came from the recent measurements
of the positron-to-electron ratio, between 1.5 and 100 GeV by the
PAMELA satellite experiment.  PAMELA found that the positron fraction
changes slope at around 10 GeV and begins to increase steadily up to
100 GeV\cite{pamela-e,pamela-e1}, confirming the tendency revealed in
some earlier experiments with lower significance and in a narrower
energy range (see Section \ref{sec:Experiments}). This rising behavior
is very different from that predicted for the secondary positrons
produced by collisions of CR nuclides and the ISM (see
eq.(\ref{eq:R_bg}))\cite{Strong98ApJ,Yin08PRD}.

\begin{figure}[t]
\begin{center}\includegraphics[width=0.8\columnwidth]
{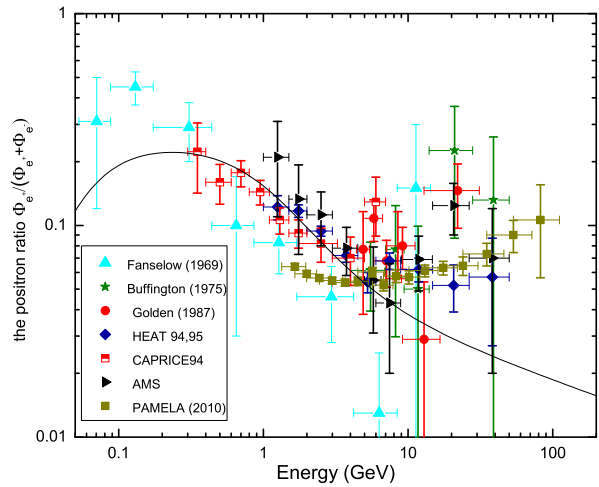}\end{center}
\caption{
\label{data-ratio}
Some measurements of the positron fraction up to an energy $\sim 100$ GeV. The data are from
Refs. 32, 49, 20, 36, 37, 43, 22. 
The solid line
is the positron fraction estimated by using
eq.(7).
}
\end{figure}

We summarize the data
in Fig.\ref{data-fermi} and Fig.\ref{data-ratio}. The latest observations call for an
additional component of electrons and positrons which is clearly
unaccounted for in the standard CR model. The lack of an anti-proton excess found
by PAMELA\cite{pamela-p}
plays a key role in understanding the origin of the
electron/positron excess.

\section{Physics of $e^\pm$ CRs: Propagation Models and Cosmic Sources}

\subsection{The $e^\pm$ background and the propagation effect}
Galactic cosmic-ray propagation is a decades-old problem
in astrophysics. The charged particles propagate diffusively in the
Galaxy due to scattering with the random magnetic
field\cite{Ginzburg64book,Strong98ApJ,Strong07Rev}.
Interactions with the interstellar medium (ISM) and the interstellar
radiation field would lead to energy losses of CRs. For heavy nuclei
and unstable nuclei, there are also fragmentation processes by
collisions with ISM and radiactive decays, respectively.
The distribution of CRs is also modified
by the convection driven by the galactic wind,
and by the re-acceleration due to the interstellar shocks.
As a result, the spectrum of the particles detected on Earth differs
significantly from those emitted from the sources. The CR propagation
equation for a particular particle species can be written in the
general form\cite{Strong98ApJ,Strong07Rev}
\begin{eqnarray}
{\partial \Psi (\vec{r},p,t) \over \partial t}&=&q(\vec{r},p,t)+
\vec{\nabla}\cdot(D_{xx}\vec{\nabla}\Psi-\vec{V}\Psi)+{\partial
\over \partial p}p^{2}D_{pp}{\partial \over \partial p}{\Psi \over p^{2}}
\nonumber\\
&&
-{\partial \over \partial p}[\dot{p}\Psi-{p \over 3}(\vec{\nabla}\cdot
\vec{V})\Psi]-{1\over \tau_{f}}\Psi-{1\over \tau_{r}}\Psi,
\label{eq:propagation}
\end{eqnarray}
where $\Psi (\vec{r},p,t)$ is the CR density per unit total particle
momentum $p$ at the position $r$, which can be expressed as
$\Psi (\vec{r},p,t)dp = 4\pi
p^{2} f (\vec{p})dp$ in terms of phase-space density $f (\vec{p})$;
$q(\vec{r}, p, t)$ is the source term including primary, spallation,
and decay contributions; $D_{xx}$ is the spatial diffusion
coefficient; $\vec{V}$ is the convection velocity; $D_{pp}$ is the
momentum diffusion coefficient that describes diffusive
reacceleration in terms of diffusion in the momentum space;
$\dot{p}$ is the momentum gain
or loss rate; $\tau_{f}$ is the timescale for loss by fragmentation;
and $\tau_{r}$ is the timescale for radioactive decay.

For primary particles such as protons and some
heavy nuclei, the source function consists of two
parts, the spatial distribution $f(\vec{r})$ and the energy spectrum
$q(p)$. The former ($f(\vec{r})$) may follow the distribution of the
possible sources of CRs, for example, the supernova remnants (SNRs).
The injection spectrum $q(p)$ is usually assumed to be a
power law (or a broken power law) function with respect to
momentum $p$. The source function of the secondary particles,
which depends on the distributions of primary CRs and the properties
of the ISM, is given by \cite{Strong98ApJ,Yin08PRD}
\begin{equation}
q(\vec{r},p,t)=\beta c \Psi_{p}(\vec{r},p,t)[\sigma_{\rm H}(p)n_{\rm  H}
(\vec{r})+\sigma_{\rm H_{\rm e}}n_{\rm H_{\rm e}}(\vec{r})],
\label{eq:secondary}
\end{equation}
where $\beta c$ is the velocity of the injecting CRs ($c$ is the speed
of light); $\Psi_{p}(\vec{r},p,t)$ is the density per unit momentum
$p$ of the primary CRs;
$\sigma_{\rm H}$ and $\sigma_{\rm He}$ are the cross sections
of producing the secondary particles from the interactions between
primaries and the H and He targets;
$n_{\rm H}$ and $n_{\rm He}$ are the interstellar hydrogen and
helium number densities, respectively. The secondary
electrons/positrons are mainly produced in the ${\rm proton}-{\rm proton}$ and ${\rm proton}-{\rm He}$
collisions, resulting in charged pions and kaons, which further decay
as $K^\pm \rightarrow \pi^\pm + \pi^{0}$, $K^\pm \rightarrow \mu^\pm +
\nu_{\mu}$, $\pi^\pm \rightarrow \mu^\pm + \nu_{\mu}$ and $\mu^\pm
\rightarrow e^\pm + \bar{\nu}_\mu+\nu_{e}$. The spatial diffusion
coefficient is usually regarded as isotropic and is described
by a function
\begin{equation}
D_{xx}=\beta D_0 (\rho /\rho_0)^{\delta},
\end{equation}
where $\rho$ is the magnetic rigidity (defined as momentum $p$
per unit charge).

In addition to spatial diffusion, scattering of CR particles on
randomly moving MHD waves leads to stochastic acceleration, which is
described in a transport equation as diffusion in momentum space
with a diffusion coefficient $D_{pp}$, which is related with the
spatial diffusion coefficient $D_{xx}$ through\cite{Seo94AA}
\begin{equation}
D_{pp}D_{xx}={4p^{2}v_{A}^{2} \over 3\delta
(4-\delta^{2})(4-\delta)w}~.
\end{equation}
Here $v_A$ is the Alfven' speed, and $w$ (may be taken as 1) is the
ratio of magnetohydrodynamic wave energy density to the magnetic field
energy density, characterizing the level of turbulence\cite{Seo94AA,Yin08PRD}.

The convection velocity $\vec{V}$, which is related to the galactic
wind, is assumed to be cylindrically symmetric and increases linearly
with height $z$ from the galactic plane\cite{Strong98ApJ}.  Besides
transporting particles, convection also causes adiabatic energy
losses of CRs as the wind speed increases away from the disk\cite{Strong07Rev}.

Practically most our knowledge of CR propagation is learned from
studying secondary CRs.
This is because the secondary production
functions can be computed  with reasonable
precision from the locally observed primary spectra,
cross sections, and interstellar gas densities.
The propagation parameters are determined by comparing the predicted
secondaries after the propagation effect with the observations.  The
typical parameters found in the CR modeling are $D_{0} \sim (3-5)
\times 10^{28} {\rm cm^{2}~s^{-1}}$ and $\delta \sim 1/3$ for
$\rho_0 \sim 1~{\rm GV}$\cite{Strong07Rev}.

In some simplified cases the propagation equation (\ref{eq:propagation})
can be solved analytically using the Green function
method\cite{Kamionkowski91PRD,Baltz98PRD}.
In most realistic cases an analytical solution is not available.
A numerical model, widely known as the GALactic PROPagation
(GALPROP; http://galprop.stanford.edu) model, was developed by Strong
and Moskalenko to solve the problem \cite{Strong98ApJ,Strong07Rev}.
The main parameters for a given GALPROP model include the CR primary
injection spectra, the spatial distribution of CR sources, the size
of the propagation region, the spatial and momentum diffusion
coefficients and their dependencies on particle rigidity. The ISM
and interstellar radiation field are adopted to calculate fragmentations
and energy losses of CRs. The parameters are tuned to reproduce the CR
spectra observed on Earth. The publicly available GALPROP code can
give relatively good descriptions of almost all kinds of CRs,
including the secondaries such as anti-proton as well as diffuse
$\gamma$ rays\cite{Strong98ApJ,Strong07Rev}.

As an example, for the Moskalenko \& Strong's model
08-005 without reacceleration\cite{Moskalenko98ApJ}, the
calculated primary electron and secondary electron and positron
background fluxes can be parameterized as follows\cite{Baltz98PRD}
\begin{eqnarray}
&&\Phi_{e^-}^{\rm bg, prim} = {0.16 \varepsilon^{-1.1}\over 1+11 \varepsilon^{0.9} + 3.2 \varepsilon^{2.15}}
({\rm GeV^{-1}~cm^{-2}~s^{-1}~sr^{-1}})\;,\nonumber\\
&& \Phi_{e^-}^{\rm bg,sec} = {0.7 \varepsilon^{0.7}\over 1 + 110 \varepsilon^{1.5} + 580 \varepsilon^{4.2}}({\rm GeV^{-1}~cm^{-2}~s^{-1}~sr^{-1}})\;,\nonumber\\
&&\Phi^{\rm bg,sec}_{e^{+}} = {4.5 \varepsilon^{0.7}\over 1 + 650 \varepsilon^{2.3} + 1500\varepsilon^{4.2}}
({\rm GeV^{-1}~cm^{-2}~s^{-1}~sr^{-1}})\;.
\end{eqnarray}
In the above the energy $\varepsilon$ is in units of GeV (i.e.,
$\varepsilon\equiv {\cal E}/{\rm 1~GeV}$). For $\varepsilon\gg 1$, we
have
\begin{equation}
\Phi_{e}=\Phi_{e^-}^{\rm bg, prim}+\Phi_{e^-}^{\rm
bg,sec}+\Phi_{e^{+}}^{\rm bg,sec}\approx \Phi_{e^-}^{\rm bg, prim}
\propto \varepsilon^{-3.25},
\end{equation}
and
\begin{equation}
{\cal R}^{\rm bg}=\Phi^{\rm bg,sec}_{e^{+}}/(\Phi_{e^-}^{\rm bg,
prim}+\Phi^{\rm bg,sec}_{e^{+}})\approx
0.06\varepsilon^{-0.25}.\label{eq:R_bg}
\end{equation}
This set of background spectra agree well with the results of some
more sophisticated numerical models\cite{Yin08PRD} derived for
$\varepsilon>10$.

The decreasing behavior of ${\cal R}^{\rm bg}$ can be understood more
straightforwardly. In the diffusion models, the diffusion coefficient can
be approximated as $D \propto \varepsilon^{\delta}$. The CRs are
assumed to be produced with a power-law spectrum, $dN/d\varepsilon
\propto \varepsilon^{-\alpha_{\rm cr}}$. The observed spectrum is then a
convolution of the source spectrum and propagation losses, giving
$\Phi_{e^{-}}^{\rm bg,prim} \propto \varepsilon^{-(\alpha_{\rm e}+
\delta)}$ for primary electrons, where $\alpha_{\rm e}$ and
$\alpha_{\rm p}$ are the source power-law indices of CR electrons
and protons, respectively. Positrons are
secondary CRs formed from CR protons, and
suffer additional propagation loses, implying $\Phi_{e^{+}}^{\rm
bg,sec} \propto \varepsilon^{-(\alpha_{\rm p}+2\delta)}$.  Then
approximately one has ${\cal R}^{\rm bg} \propto E^{ \alpha_{\rm
e}-\alpha_{\rm p}-\delta}$ as long as $\Phi_{e^{-}}^{\rm
bg,prim}>\Phi_{e^{+}}^{\rm bg,sec}+\Phi_{e^{-}}^{\rm bg,sec}$.
Usually one adopts $\alpha_{\rm e} \approx \alpha_{\rm p}$, so the
standard model then predicts a CR positron-to-electron ratio that
decreases with energy\cite{Serpico09,Shaviv09,Piran09}, in agreement
with the tendency reported in eq.(\ref{eq:R_bg}).

In the GALPROP code, there are several main approximations: a) The
distribution of sources is continuous.  b) All CRs are assumed to be
from one single type of astrophysical sources, namely the
supernova remnants (SNRs).  c) The propagation of CRs
is isotropic. Considering anisotropic propagation models can help to
solve the problem of soft gradient in the radial dependence of the
$\gamma-$ray flux, and can explain the large bulge-over-disk ratio
in positron annihilation as observed by INTEGRAL. However, as shown
in Fig.4 of Ref.\refcite{Gebauer09}, the anisotropic propagation
model gives a similar CR electron spectrum as the isotropic
model in the energy range $>10$ GeV. We therefore will not discuss the
anisotropic propagation effect in the rest of the review. The change
of the first two assumptions can give rise to interesting CR spectral
signals, which we will come back to discuss later.

\subsection{Sources of $e^{\pm}$ cosmic rays}
In general, there are five types of high energy electron/positron
production processes:

\begin{itemize}
\item Acceleration of electrons in shocks or in magnetic
turbulence. In strong shocks the highest electron energy
is governed by the balance between energy gain (via shock
acceleration) and
loss (via synchrotron radiation in the magnetic field) and can be
estimated as ${\cal E}_{\rm e,max}\sim 20~{\rm TeV}~\beta_{\rm
sh}\Gamma {B_{\rm sh}}^{-1/2}$, where $B_{\rm sh}$ is the strength of
the magnetic field in the shock region, $\beta_{\rm sh}$ is the
velocity of the shock in units of speed of light $c$ and $\Gamma$
is the bulk Lorentz factor of the shock region.

\item Cosmic rays accelerated at shocks produce secondary
electrons/positrons inside the source through hadronic interactions.
These interactions  usually
lead to production of charged pions or kaons, which decay to produce
electrons and positrons. Besides the ${\rm proton}-{\rm proton}$ and ${\rm proton}-{\rm He}$
interactions discussed following eq.(3), some other
processes include interactions between hadrons and photons, e.g.
${\rm proton}+\gamma_{\rm bg}\rightarrow \Delta^{+} \rightarrow
{\rm neutron}+\pi^{+}\rightarrow {\rm neutron}+e^{+}+\nu_{\rm e}+\bar{\nu}_{\mu}+\nu_{\mu}$
(${\rm proton}-\gamma$ process)
and ${\rm particle}+\gamma_{\rm bg}\rightarrow {\rm particle}
+e^{+}+e^{-1}$, for
which the energy thresholds are ${\cal E}_{\rm proton}{\cal E}_{\gamma_{\rm
bg}}\approx (0.3~{\rm GeV})^{2}$ and ${\cal E}_{\rm particle} {\cal
E}_{\gamma_{\rm bg}} \approx m_{\rm e}m_{\rm particle}c^{4}$,
respectively.

\item Annihilation of high energy photons with  background
photons, i.e., photon-photon pair production, $\gamma_{\rm high} +
\gamma_{\rm bg} \rightarrow e^{+}+e^{-}$. The energy threshold is
${\cal E}_{\gamma_{\rm high}}{\cal E}_{\gamma_{\rm bg}} \approx
2(m_{\rm e}c^{2})^{2}$ \cite{Jauch76book}.

\item Pair production in ultra-strong magnetic fields, i.e.,
photon-magnetic field pair production, $\gamma + B \rightarrow
e^{+}+e^{-}$. The magnetic field $B$ should satisfy $B\geq 6\times
10^{13}~{\rm G}~({\cal E}_{\gamma}/1~{\rm MeV})^{-1} \sin
\vartheta^{-1}$, where $\vartheta$ is the angle between the photon and
the magnetic field line\cite{Erber66RMP}. Such a strong magnetic
field is expected in the vicinity of pulsars.

\item Annihilation or decay of dark matter
particles. Different from the above four processes, such kind of
scenarios, although well motivated, are highly speculative and lack
of solid experimental evidence.
\end{itemize}

\subsection{Age and distance of the sources of $e^\pm$ excess}
\label{sec:age-dist}

The above five processes can take place in various astrophysical
sources. However, it is difficult to pin down the exact source(s)
of the observed $e^\pm$ excess based
on observations. This is because before reaching Earth the charged
particles have been deflected by the interstellar magnetic fields.
One can see this effect by estimating the Larmor radius $R_{\rm L} \sim
10^{15}~{\rm cm}~({\cal E}/1~{\rm TeV})(B_{\rm IG}/3{\rm \mu
G})^{-1}$, which is much smaller than the distance of the CR sources to
Earth, where $B_{\rm IG} \sim 3\mu{\rm G}$ is the strength of the
interstellar magnetic field, and $u_{\rm cmb}$.
Nonetheless, one can reliably estimate
the age and the distance of these sources in the following way.
For an electron with energy ${\cal E}_{\rm e}$, the total power of
synchrotron and inverse Compton radiation in
the Thomson regime is $P_{\rm r}=\ell_{0}({\cal E}/m_{\rm
e}c^{2})^{2}$, where $\ell_{0}\equiv 4\sigma_{\rm T} c u_{\rm tot}/3$.
The characteristic cooling timescale can be estimated as $\sim
{\cal E}/P_{\rm r}$, i.e.,
\begin{equation}
\tau_{\rm rad}\approx {3m_{\rm e}^{2}c^{3} \over 4\sigma_{\rm T}u_{\rm
tot}{\cal E}}\approx 10^{15}~{\rm s}~({u_{\rm tot}\over
1~{\rm eV~cm^{-3}}})^{-1}({{\cal E} \over 10~{\rm GeV}})^{-1},
\end{equation}
where $u_{\rm tot}=u_{\rm B}+u_{\rm cmb}+u_{\rm dust}+u_{\rm
star}$, $u_{\rm B}=B_{\rm IG}^{2}/8\pi$ is the magnetic field energy
density, $u_{\rm dust}$ and
$u_{\rm star}$ are the photon energy densities of cosmological
microwave background, dust emission, the star emission, respectively.
The travel distance of these CR electrons/positrons cannot exceed the
characteristic CR diffusion radius
\begin{eqnarray}
R_{\rm diff} &\sim & 2\sqrt{D_0({\cal E}/10~{\rm
GeV})^{\delta} \cdot c \tau_{\rm rad}}\nonumber\\ &\sim & 4~{\rm
kpc}~({D_0 \over 10^{28}~{\rm cm^{2}~s^{-1}}})^{1\over 2}({u_{\rm
tot}\over 1~{\rm eV~cm^{-3}}})^{-{1\over 2}}({{\cal E} \over
10~{\rm GeV}})^{\delta-1\over 2}.\label{eq:R_diff}
\end{eqnarray}
For $\delta\sim 1/3$, we find that for the observed
electrons/positrons at $\sim 1$ TeV,
{\it the source should have an age $\leq
\tau_{\rm rad} \sim 10^{13}$ s and should be within a distance $\leq
R_{\rm diff} \sim 1$ kpc.}
As a result, statistical fluctuations in the
injection spectrum and spatial distribution of nearby sources
within $R_{\rm diff}$ may cause significant deviations from
the naive predictions in the conventional homogeneous and steady
state scenario\cite{Aharonian95AA,Strong01ICRC,Kobayashi04ApJ}.
Since $R_{\rm diff}$ is anti-correlated to ${\cal E}$, the
deviations are more prominent at high energies.

In the following three sections, we will discuss three types of models
that have been proposed to interpret the observed $e^\pm$ excess,
namely, the modified SNR models for the $e^\pm$ background, new
astrophysical sources, and dark matter.

\section{Scenario I: Modified SNR Models for the $e^\pm$ Background}
Supernova remnants (SNRs) are the canonical
sources of CRs\cite{Hillas05}. The kinetic
energy of the SNRs is found to be comparable to that needed to
accelerate Galactic CRs. The more direct evidence is the
detection of the photons up to $\geq 10$ TeV from
some SNRs, which are consistent with $\pi^0$ decay from hadronic
interactions\cite{Aharonian06AA}. SNRs have been traditionally regarded
as the dominant source to contribute to the observed $e^\pm$
background. This model can naturally account for a smooth, gradually
softening $e^\pm$ spectrum at high energies. The excess feature discussed
above is not expected from the simplest model. In this section we discuss
whether the excess is simply due to some factors that have been
ignored in the conventional SNR models.

\subsection{Inhomogeneity of SNRs?}
As already mentioned, in the widely adopted code GALPROP the
distribution of SNRs is assumed to be homogenous. Such an approximation
would not affect the results of ion CRs significantly.  For electrons,
on the other hand, due to the significant cooling caused by magnetic
fields and background photons, only the electrons generated within
a distance $\sim 1$ kpc can reach Earth\cite{Aharonian95AA,Strong01ICRC}
(see also Section \ref{sec:age-dist}). Within such a small volume, the
SNR source distribution can be very inhomogeneous. Releasing the
homogeneity assumption could therefore lead to significant
modification of the detected $e^\pm$ spectrum.

\begin{figure}[t]
\begin{center}\includegraphics[width=0.7\columnwidth
]{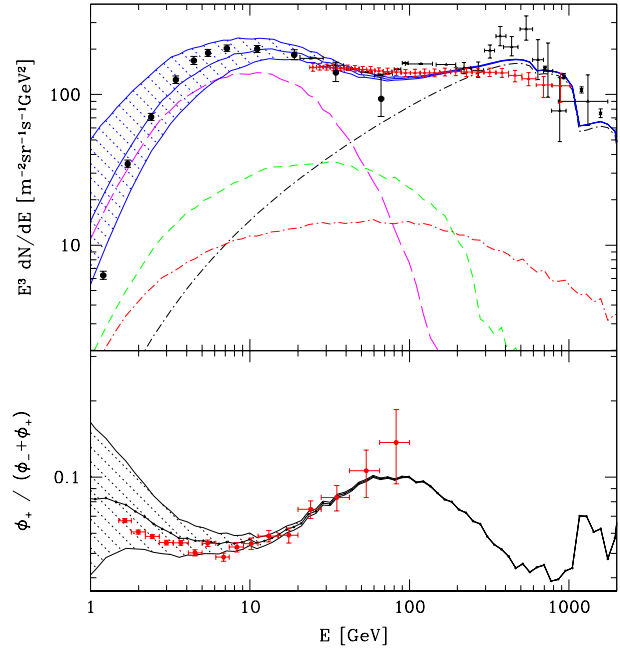}\end{center}
\caption{
\label{fig:SNR-Shaviv}
Predictions of the inhomogeneous SNR models as compared
with data (from Ref.66). {\it Top Panel}: The
expected electron and positron spectra -- Primary arm electrons
(long dashed purple), primary disk electrons with nearby sources
excluded (short dashed green), nearby SNRs (dot-dashed black),
secondary positrons (dot-dashed red), and their sum (blue). The
hatched region describes the solar modulation range (0.2-1.2 GeV).
{\it Bottom Panel}: Model results and the measured PAMELA
data for the positron fraction.  The shaded region is
variability expected from solar modulation effects (from Ref.66).
}
\end{figure}

In the Milky Way, star formation is concentrated in spiral
arms\cite{Lacey01ApJ}. The nearest spiral arm, the Sagittarious-Carina
arm, is about 1 kpc away. The inhomogeneity effect of SR source
distribution in such a distance scale is therefore very important
to shape the observed electron/positron CR spectrum\cite{Shaviv09}.
Since $R_{\rm diff}$ is energy dependent (eq.(\ref{eq:R_diff})),
an energy-dependent feature would appear. Low energy electrons
($\sim $ 1 GeV) can easily reach us from a larger radius where
inhomogeneity is more averaged out. On the other hand, electrons
with higher energies ($\geq 10$ GeV) can arrive Earth only from
very nearby sources since they would otherwise have cooled via
synchrotron and inverse-Compton radiation before reaching Earth.
These very nearby sources would present a bump in the $e^\pm$
spectrum in the $>100$ GeV range, as observed by ATIC and Fermi.
Pairs formed in the local vicinity through the proton/ISM interactions can
reach the solar system also at high energies.
As shown in Fig.\ref{fig:SNR-Shaviv}, both the PAMELA-observed
increase of the positron-to-electron ratio between 1 and 100 GeV
and the ATIC-observed $e^\pm$ excess around $\sim$ 600 GeV
can be attributed to the inhomogeneity of SNRs
\cite{Shaviv09,Piran09}.

One interesting prediction of the inhomogeneous SNR model is that the
positron-to-electron ratio should start to drop at energy ${\cal
E}\sim$ 100 GeV, just above the present PAMELA measurement (see
Fig.\ref{fig:SNR-Shaviv}). It should reach a minimum around the ``ATIC
peak", where it should start to rise again. This is because the nearby
SNRs mostly contribute to primary electrons near the ``ATIC peak'',
without significantly increasing secondary positrons. Such a
prediction is in stark contrast to the case where the $>100$ GeV excess
is due to a primary source of pairs (e.g. either the pulsar model
or the dark matter models), in which the positron fraction
is expected to keep rising at a few hundreds GeV until reaching
$\sim 50\%$. The positron fraction behavior above 100 GeV is the
smoking gun to test this inhomogeneous SNR model (and the pair
production models).

\subsection{Acceleration of the secondary $e^\pm$ in SNRs?}
\begin{figure}[t]
\begin{center}\includegraphics[width=0.8\columnwidth
]{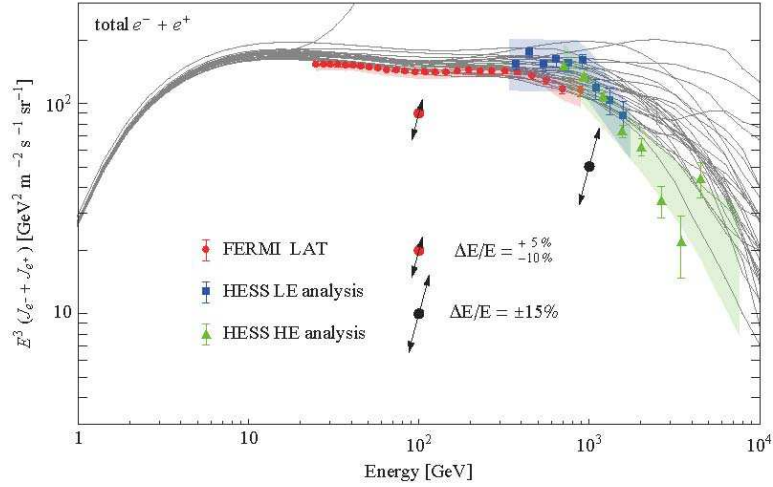}\end{center}
\caption{
\label{fig:Ahlers09PRD}
The spectrum of the total cosmic ray electrons and positrons
expected in the secondary particle acceleration model
(The figure is from Ref.80).
}
\end{figure}

In the standard model, primary cosmic rays are accelerated in SNRs
while the secondary $e^\pm$ are not.  In reality the secondary
production also takes place in the same region where cosmic rays are being
accelerated. These secondary $e^{\pm}$ participate in the acceleration
process and turn out to have a flatter spectrum relative to the
primaries at high energies. This could be responsible, after
propagation in the Galaxy, for the observed PAMELA
anomaly\cite{Blasi09PRL,Ahlers09PRD}. As found in solving the transport
equation, the downstream equilibrium spectrum of secondary electrons
produced in the acceleration region has two terms.  One term traces
the spectrum of the high energy primary CRs, i.e., $\propto {\cal
E}^{-\kappa}$. The other term has a spectrum $\propto {\cal
E}^{-\kappa+\delta}$, where the factor ${\cal E}^{\delta}$ arises
from the momentum dependence of the diffusion
coefficient\cite{Blasi09PRL,Ahlers09PRD}. As long as the second term
dominates, the spectrum gets flattened.  This effect was noted
earlier as a general expectation for the secondary-to-primary ratio
in the presence of continuous Fermi
acceleration\cite{Eichler80ApJ,Cowsik80ApJ}.  By simulating the
spatial and temporal distributions of SNRs in the Galaxy
according to their known distribution statistics, the $e^{+}+e^{-}$
energy spectrum measured by Fermi LAT can be interpreted within
such a secondary $e^\pm$ acceleration model\cite{Ahlers09PRD} (see
Fig.\ref{fig:Ahlers09PRD}). A large portion of the model parameter
space seems to lead to overprediction of TeV flux as compared with
the H.E.S.S Observations. Considering the large systematic
uncertainties of H.E.S.S. and large uncertainties of the model
parameters, the model can make a case to account for both the
$e^\pm$ excess and the positron fraction excess. Future high
quality TeV data can give a more reliable test to this model.

The secondary particle acceleration model has several interesting
predictions. In particular, all other secondary particles (including
baryonic particles) are also accelerated. As a result,
the antiproton to proton ratio should
also increase with energy. This is a clean criterion to differentiate
this model from some other astrophysical models for positron fraction
excess (e.g. pulsar models). Some dark matter models however share the
same prediction, and therefore cannot be differentiated based on this
criterion.  A similar effect applies to secondary nuclei such
as titanium and boron \cite{Mertsch09PRL}. These predictions can be soon tested with the
data from PAMELA and the forthcoming AMS-2 mission. Finally, since
this model invokes strong hadronic interactions, one would naturally
expect strong hadronic neutrino signals from nearby SNRs, so that
IceCube has good prospects to detect them.

This model requires that secondary positrons can escape the shock
region to reach Earth. For young SNRs, magnetic field amplification
by the shock wave is very effective, so that synchrotron cooling of
$e^\pm$ is significant. High energy $e^\pm$ may not be able to
escape from the acceleration site. As a result, the SNRs invoked
in this model should be old enough.

\subsection{$e^\pm$ pair production in the photon-proton interaction in young SNRs?}
\label{sec:youngSNR}

Despite the cooling constraint discussed above,
young SNRs have been argued to be an attractive candidate to solve the
excess puzzle. In young SNRs, interactions of newly formed CR nuclei
and background radiation may produce energetic $e^{\pm}$ pairs, which
would change the CR spectrum at the source. Considering a photon field
at the source peaking in the optical/infrared band (${\cal
E}_{\gamma_{\rm bg}}\sim$ eV), pair production occurs at $\sim
(m_{\rm e}c^{2}/{\cal E}_{\gamma_{\rm bg}})m_{\rm particle}c^{2}$,
which is $\sim 1$ PeV for protons and $\sim$ a few PeV for
helium\cite{Blum70PRD}. These correspond to the ``knee'' of the
cosmic CR spectrum. In the rest frame of the CR
nuclei, the secondary electrons/positrons have energies around MeV,
which turn out to be about TeV in the observer's frame. It is striking
to note that this is just the energy range where the $e^\pm$ excesses
were observed by ATIC, Fermi, and H.E.S.S.  Furthermore, the
energy density of the excess electrons observed by ATIC is about
$u_{e^\pm}\sim 3\times 10^{-5} ~{\rm eV~ cm^{-3}}$ between 0.1 and
1 TeV, which is of the order of the energy density due to CR energy loss
assuming a spectral break from 2.7 to 3.1 at energy $\sim 1$ PeV.
These coincidences are the motivations of the photon-CR interaction
model\cite{Hu09ApJL}. Indeed the
interaction between CR nuclei and background photons may naturally
bridge the ``knee'' of the CR spectrum and the $e^\pm$ excess,
provided that the optical background photons have a thermal
distribution with a temperature $T_{\rm opt}\sim 5000-7000$
K\cite{Hu09ApJL}. Such a dense background photon field is
possible in the vicinity of young SNRs.

One potential problem not addressed by Ref.\refcite{Hu09ApJL} is
whether the resulting pairs can escape from the source. As discussed
above, young SNRs tend to have strong magnetic fields. Together with
the very dense background photons invoked to produce $e^\pm$, strong
cooling of the produced secondary $e^\pm$ is expected, which would
lead to the production of strong infrared/optical synchrotron radiation
and GeV-TeV inverse Compton radiation. One needs to check whether
the escaped $e^\pm$ can still reach the desired energy to account for
the observed excess.

\subsection{Klein-Nishina suppression of electron/positron cooling?}
An electron or positron moving in a dense soft photon bath would lose
energy via inverse Compton (IC) scattering\cite{Blumenthal70RMP} and
produce a high-energy IC component at
${\cal E}_{\rm ic} \approx {2\gamma_{\rm e}^{2} {\cal
E}_{\gamma_{\rm bg}}\over 1+g}$, where $g \equiv {\gamma_{\rm e}}
{\cal E}_{\gamma_{\rm bg}}/m_{\rm e}c^2$ and $\gamma_{\rm e}\equiv{\cal E}/m_{\rm e}c^{2}$ is the Lorentz factor
of the electron.  In the Thompson regime, $g
\ll 1$, so ${\cal E}_{\rm ic} \approx 2 {\gamma_{\rm e}^{2} {\cal
E}_{\gamma_{\rm bg}}}$.  In the extreme Klein-Nishina regime, $g \gg
1$, one has ${\cal E}_{\rm ic} \approx \gamma_{\rm e} m_{\rm e}c^2$.
The scattering cross section in the Klein-Nishina regime is also greatly
reduced, so that the electron/positron cooling rate drops
correspondingly.
The strength of the drop is proportional to the
energy density of the seed photons.
One may define the critical Klein-Nishina energy of electrons ${\cal
E}_{\rm KN} \sim 0.3m_{\rm e}^{2} c^{4}/{\cal E}_{\rm bg}$,
above which is the Klein-Nishina effect becomes
significant\cite{Schlickeiser09AA}.
This corresponds to $\sim 44$ GeV for a typical B star and $\sim
180$ GeV for typical G-K stars. So given the galactic diffuse
optical background from stars, the effect takes place in the
energy range where electron/positron excess is observed to
appear\cite{Schlickeiser09AA,Stawarz09ApJ}.
Associated with the drop in energy loss rate in the Klein-Nishina regime is
a gradual hardening of the cosmic ray $e^\pm$ spectrum.
The spectral structure in the FERMI and H.E.S.S. data may
be reproduced in this way\cite{Schlickeiser09AA,Stawarz09ApJ}
provided that: (1) The energy density of the seed photons is
significantly higher than that of the magnetic field (so that
IC cooling is more significant than synchrotron cooling); (2) the
star-to-dust emission ratio $u_{\rm star}/u_{\rm dust}$ is large
enough\cite{Stawarz09ApJ}, so that the dominant IC cooling is with
respect to star optical emission. This requirement is needed since
IC off the dust emission in the far-infrared range is still in the
Thomson regime even for relatively energetic electrons. Such a cooling
effect would compete with the Klein-Nishina cooling effect if $u_{\rm dust}$ is
not suppressed, which would smear the $e^\pm$ spectral feature.
This Klein-Nishina-related feature was not noticed previously in the
GALPROP calculations\cite{Strong98ApJ} since a small $u_{\rm star}/u_{\rm dust}$ was adopted.

The Klein-Nishina effect affects both primary electrons and secondary
positrons. Besides accounting for the excess in the total $e^\pm$
spectrum, it can affect the
positron-to-electron ratio and may be able to reproduce the PAMELA
result\cite{Stawarz09ApJ}. The $u_{\rm star}$ and the number density
of the interstellar medium are however required to be about 100 times
larger than the average values expected in the Galactic disk. Such
extreme parameters may apply to the vicinities of young supernova
remnants. A caveat is again high-energy $e^{\pm}$ may not be able to
overcome strong cooling and escape from such sources
(see \S\ref{sec:youngSNR}).

\section{Scenario II: New Astrophysical Sources}
In the previous section, it is assumed that SNRs are the sole source of
cosmic ray electrons/positrons. Such an assumption may well be wrong since
other astrophysical objects can also produce
high energy electron/positron cosmic rays. As early as in 1970s, pulsars
had been suggested as the source of the CR electrons/positrons\cite{Shen70}.
Some widely discussed objects include Vela and
Geminga\cite{Shen70,Aharonian95AA,Yuksel09PRL,Hooper09JCAP}.
Some other nearby cosmic ray accelerators, such as microquasars or an
ancient gamma-ray burst, are also possible candidates. In this section
we discuss these alternative astrophysical sources.

\subsection{Pulsars}
Pulsars are strongly magnetized, rapidly rotating neutron stars that
are powered by spin down and emit broad-band pulsed electromagnetic
radiation\cite{Manchester77}.
They are important particle accelerators. Direct evidence comes from
the recent Fermi discovery of nearly 50 high confidence gamma-ray
pulsars with emission above 100 MeV\cite{Abdo09Sci,Abdo09ApJ}.
Three accelerator sites have been discussed in the literature.
1. Polar cap models invoke a charge-depleted gap near the magnetic
polar cap region\cite{Ruderman75ApJ,Usov96ApJ,Arons83ApJ} in which
primary particles are accelerated and high energy $\gamma$-rays are
radiated via curvature radiation of inverse Compton scattering. The
high energy $\gamma$-rays interact with strong magnetic
fields near the pole and produce electron-positron pairs. These
secondary pairs cool via synchrotron radiation or resonant inverse
Compton scattering and produce higher generation $\gamma$-rays and
pairs, leading to a photon-pair cascade\cite{Daugherty96ApJ,Zhang00ApJ}.
2. Outer gap models\cite{Cheng86ApJ,Romani96ApJ,Zhang97ApJ}
invoke a charge depleted gap developed beyond the ``null'' charge
surface but enclosed by the last open field line.
Particles are accelerated to high energies in the intense electric field
parallel to the magnetic field lines inside the outer gap, and
$\gamma$-rays are produced via inverse Compton scattering or curvature
radiation. The $\gamma$-rays interact with other softer photons in
the magnetosphere to produce $e^\pm$ pairs.
3. Slot gap models\cite{Arons79ApJ,Muslimov03ApJ} invoke an elongated
gap near the last open field line, the electric field inside which is
immune from pair screening thanks to its favorable geometry. Particles
can be accelerated from the surface all the way to the light cylinder.
Almost the entire electric potential can be utilized for
acceleration. Outer gaps cannot exist in old pulsars (e.g.
$t>10^{7}$ yr), while polar gaps and slot gaps can exist in
all active pulsars. Recent Fermi observations disfavor the polar cap
origin of most $\gamma$-rays, although both the outer gap and the slot
gap models are still reasonable candidates for pulsar $\gamma$-ray
emission, and hence, for energetic particle acceleration.
Outside the magnetosphere, pairs can be re-accelerated in shocks as the
pulsar wind interacts with the ambient medium. For young pulsar
systems, a bright pulsar wind nebula would exist inside the supernova
remnant, which radiate in a broad band from X-rays up to TeV energies.

To estimate the CR electron/positron production from pulsars, one needs
to know the total energy budget of a pulsar and the fraction of this
energy that is converted into pairs. For the first purpose we adopt the
classical rotating magnetic dipole model
\cite{Pacini67Nature,Gold68Nature,Ostriker69ApJ}, in which the pulsar
loses its rotational energy through magnetic dipole radiation.
The spin-down power is estimated as
\begin{equation}
L_{\rm dip}=I\Omega \dot{\Omega} \simeq 2.6\times 10^{40}~{\rm
erg~s^{-1}}~B_{p,12}^2R_{\rm
s,6}^6\Omega_{0,3}^{4}(1+t/\tau_{0})^{-2}, \label{eq:L_dip}
\end{equation}
where $B_{p}$ is the polar cap dipolar magnetic field strength of the
pulsar, $R_{\rm s}$ is the pulsar radius,
$\Omega_0$ is the initial angular frequency
of rotation (the current angular frequency of the rotation is
$\Omega=2\pi/P=\Omega_{0}(1+t/\tau_{0})^{-1/2}$), $\tau_{0}=1.6\times
10^{10} B_{p,12}^{-2} \Omega_{0,3}^{-2}I_{45}R_{\rm s,6}^{-6}$ sec
is the initial spin-down timescale of the pulsar, $I\sim 10^{45}~{\rm
g~cm^2}$ is the typical moment of inertia of the
pulsar\cite{Pacini67Nature}. Here the convention $Q_x=Q/10^x$
has been adopted in cgs units. The polar cap dipolar magnetic field
strength at the surface is $B_p = 6.4\times
10^{19}\sqrt{P\dot{P}}$ Gauss.
Following Refs.\refcite{Malyshev09PRD,Profumo09}
we denote $f_{e^{\pm}}$ as the fraction of the rotational energy
that is deposited in the cosmic ray $e^{\pm}$ pairs.
The energy output rate of $e^{\pm}$ pairs can be then estimated
as $\dot{E}_{\rm e^{\pm},out}=f_{e^{\pm}} I\Omega \dot{\Omega}$. It is
usually suggested that the mature pulsars (with ages $\sim 10^{5}$ years
and without associated SNRs) rather than the young pulsars can
contribute to the observed $e^{\pm}$ pair flux because the produced
pairs are no longer trapped and lose energy in the nebula. The
total energy output of the $e^{\pm}$ pairs for a mature pulsar can be
estimated as
\begin{equation}
{E}_{\rm e,out}={f_{e^{\pm}}\over 1+x_0} {I \Omega_0^{2} \over 2},
\end{equation}
where $x_0 \sim 10^{5}~{\rm years}/\tau_0$. The parameter $f_{e^{\pm}}$
may be time-dependent and is not well determined. A quantitative
discussion of plausible values was recently presented in
Ref. \refcite{Malyshev09PRD} (Some more model-dependent estimates of
the $e^{\pm}$ pair production rate can be found in
Refs.\refcite{Harding87ApJ,Chi96ApJ,Zhang01AA}).  In the context of a
standard model of pulsar wind nebulae, a reasonable range for
$f_{e^{\pm}}$ falls between $1\%$ and $30\%$. For typical parameters
$(I_{45},~\Omega_0,~x_0) \sim (1,~300,~10)$, we have ${E}_{\rm
e,out}\sim 5\times 10^{47}(f_{e^{\pm}}/0.1)$ ergs.  To account for the
observed ATIC electron/positron excess by a single pulsar, the pulsar
needs to be within a radius $r\sim 0.4~{\rm kpc}~({E}_{\rm e,out}/5\times
10^{47}~{\rm erg})^{1/3}(u_{e^{\pm}}/3\times 10^{-5}~{\rm
eV~cm^{-3}})^{-1/3}$. Considering that only a fraction of
electrons/positrons can reach an energy above 100 GeV, the source should
be even closer. Indeed there are a few nearby intermediate age pulsars
within 0.3 kpc (e.g. Geminga at 0.16 kpc, PSR 0656+14 at 0.29 kpc, and
Vela at 0.29 kpc). For a complete pulsar catalog with distance
information, see
http://www.atnf.csiro.au/research/pulsar/psrcat/.

Once the total energy output is specified, the injected
electron-positron spectrum is fully defined by its spectral shape.
The pairs generated from various gaps usually are not energetic
enough, and they usually do not form a simple power law spectrum.
In order to be relevant to the CR $e^\pm$ excess, one needs to assume
that these pairs are re-accelerated in pulsar wind nebula shocks.
It is well known that the $e^{\pm}$ escaping pulsar wind cannot be
accelerated to arbitrarily high energies. Depending on the pulsar
environment, a cut-off is expected, usually in the energy range
around TeV. The precise position of the cutoff $\gamma_{\rm
e^{\pm},max}$ is rather uncertain, and it critically depends on the
pulsar age and on the magnetic field strength in the pulsar wind
nebula\cite{Busching08}.

The pulsar pair spectrum is usually assumed to take the form
\begin{equation}
{dN_{\rm e^{\pm}}\over d\gamma_{\rm e^{\pm}}}\propto \gamma_{\rm
e^{\pm}}^{-\kappa}e^{-\gamma_{\rm e^{\pm}}/\gamma_{\rm e^{\pm},max}}.
\end{equation}
For a point source with a power-law spectrum ${dN_{\rm e^{\pm}}\over
d\gamma_{\rm e^{\pm}}}\propto \gamma_{\rm e^{\pm}}^{-\kappa}$, the
energy distribution of the pairs reaching Earth can be
conveniently handled as\cite{Atoyan95PRD}
\begin{equation}
n(R,t,\gamma_{\rm e^{\pm}})={dN_{\rm e^{\pm}}/d\gamma_{\rm e^{\pm}}
\over \pi^{3/2} r^{3}}{[r/R_{\rm diff}(t,\gamma_{\rm
e^{\pm}})]^{3}e^{-[r/R_{\rm diff}(t,\gamma_{\rm e^{\pm}})]^{2}} \over
(1-\ell_{0} t \gamma_{\rm e^{\pm}})^{2-\alpha}},
\end{equation}
where the spectrum has a cut off at $\gamma_c=1/(\ell_{0} t)$. The energy
loss rate and age then set a maximal energy of particles that reach
Earth today, with a diffusion radius\cite{Atoyan95PRD}
\begin{equation}
R_{\rm diff}(t,\gamma_{\rm e^{\pm}})\approx 2\{ {\cal D}(\gamma_{\rm
e^{\pm}})ct[1-(1-\gamma_{\rm e^{\pm}}/\gamma_{\rm
c})^{1-\delta}]/[(1-\delta)\gamma_{\rm e^{\pm}}/\gamma_{\rm
c}]\}^{1/2},
\end{equation}
where ${\cal D}(\gamma_{\rm e^{\pm}})\approx {\cal
D}_{0}(1+\gamma_{\rm e^{\pm}}/\gamma_*)^{\delta}$ and $\gamma_*
\approx 6\times 10^{3}$.

\begin{figure}[t]
\begin{center}\includegraphics[width=0.48\columnwidth
]{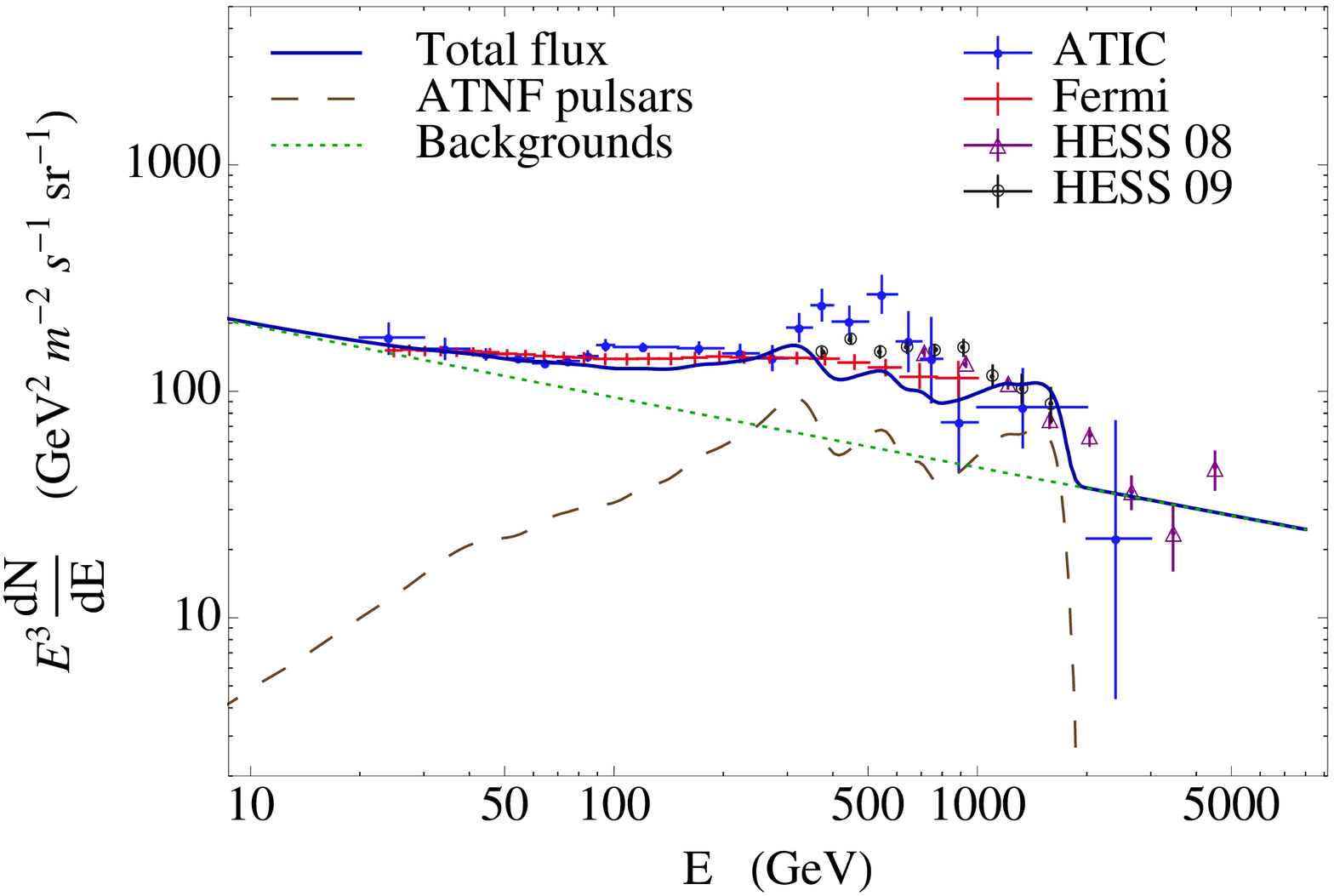}\includegraphics[width=0.48\columnwidth
]{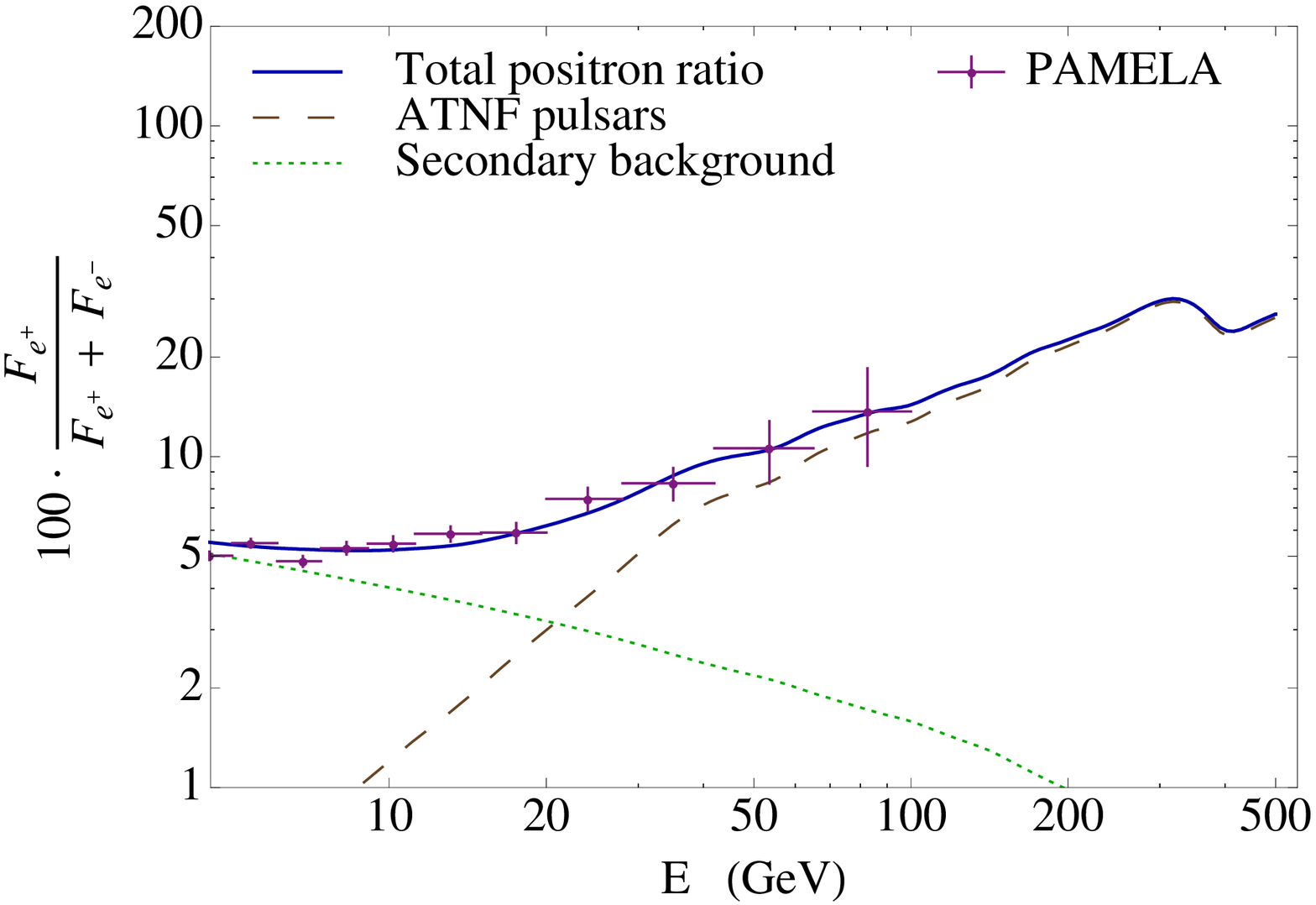}\end{center}
\caption{
\label{fig:pulsar}
The spectrum of cosmic ray electrons and positrons expected in
the pulsar model (from Ref.109).
}
\end{figure}

Following the above procedure, several authors have calculated the contribution of one or more nearby
pulsars to the observed $e^\pm$
spectrum\cite{Malyshev09PRD,Profumo09,Pohl09PRD}.
By picking a few nearby known pulsars and employing some suggested
energy output models, one is able to reproduce the spectral features
and the intensities of the reported $e^\pm$ excess,
with reasonable values for the $e^{\pm}$ output efficiency
(see Fig.\ref{fig:pulsar} for illustration).  The advantage of this
model is the lack of an anti-proton excess, in agreement with the PAMELA
observation\cite{pamela-p}.  Since
electrons and positrons are produced in pairs in the pulsar models,
the positron fraction would achieve 50\% at high energies in the
pulsar model. Another prediction of the pulsar model for
the $e^\pm$ excess is that at higher energies, the electron/positron
flux measured at Earth will be dominated by a few nearby
pulsars, and therefore the spectrum would host wiggle-like
features\cite{Kobayashi99ICRC,Malyshev09PRD,Profumo09}. The presence of such
features at high energies would strongly suggest a pulsar origin of
the anomalous contribution to the electron/positron fluxes. The
hardening of the CR electron spectrum detected by H.E.S.S at an energy
$\sim 4$ TeV, if confirmed in the future, may be taken as a support
to the pulsar model, although the inhomogeneous distribution of SNRs
may also produce the wiggle-like structure in the electron/positron
spectrum at TeV energies. The differentiation could be made by
evaluating the corresponding positron-to-electron ratio, which
should be small for the SNR model (since secondary positrons are
much rarer than primary electrons), but large for the pulsar models
(electrons and positrons are produced in pairs).

\subsection{Microqusars}
Microquasars are a sub-group of accretion-powered X-ray binaries
that process a relativistic jet. The companion can be either a high
mass or a low mass star. The accretor itself is likely a rapidly
spinning black hole, although a neutron star accretor is not ruled out.
These objects display in miniature some of the properties of quasars,
accreting supermassive black holes with much higher luminosities,
and hence, carry the name of ``microquasar''\cite{Mirabel99ARAA}.

Although their physical origin is still
unclear\cite{Bosch-Ramon08IJMPD},
microquasars can be important sources of
high energy cosmic rays\cite{Heinz02AA,Fender05MN}.
Gamma-ray flares with energy above 100 MeV have been detected
from microquasars LS I +61 303\cite{Albert06Science} and
Cygnus X-3\cite{Tavani09Nature}, suggesting that they are powerful
lepton accelerators. TeV electrons/positrons may be accelerated
in the internal or termination shocks. More pairs can be produced
as TeV gamma-rays attenuate with the UV/optical photons from the
companion star.  The net energy output of microquasars in the Galaxy is
uncertain. An optimistic estimate suggests that up to $5-10\%$ of the
total CR luminosity can be contributed by such kind of
objects\cite{Heinz02AA,Fender05MN}. If correct, the energy output of
microquasars is high enough to account for the detected CR $e^{\pm}$
excess, as noted in Refs.\refcite{atic,pamela-e}. However,
there are not many known nearby microquasars\cite{Guessoum06AA}.
The closest candidate is Cen X-4
at $\sim 1.2$ kpc.
There is no detailed calculation of this model in the literature.

\subsection{Gamma ray Bursts}
Gamma-ray bursts (GRBs) are brief, intense flashes of soft ($0.01-1$
MeV) $\gamma$-rays in the sky, which are related to collapses of some
massive stars or mergers of two compact stellar objects (two neutron
stars or a neutron star and a black hole) \cite{Piran04RMP}.
They were serendipitously discovered by
Vela-satellites in late 1960s\cite{Klebesadel73ApJ}, and now are
routinely detected by gamma-ray detectors and followed up by various
telescopes in broad band (from radio to TeV).
The event rate of nearby high-luminosity GRBs (those typical GRBs
detected at cosmological distances) is roughly \cite{Guetta05ApJ}
$0.5-1 {\rm Gpc^{-3}yr^{-1}}$, which corresponds to
about 0.025-0.05 GRB per Myr per galaxy\cite{Zhang04IJMPA}. Counting
for those GRBs that do not beam towards us (with large uncertainty
in the beaming correction factor)\cite{Frail01ApJ} - which are
relevant since off-beam GRBs can also contribute to the diffuse
$e^\pm$ background -  the rate can be
as high as 1 GRB per $10^5-10^6$ yr per galaxy. Nearby low-luminosity
GRBs can have a higher observed event rate (by a factor of
200-500)\cite{Coward05MN,Liang07ApJ}. Even if their collimation angles are
typically wider, the collimation-corrected event rate could be still
higher than that of high-luminosity GRBs.
As a result, it is possible that an ancient GRB in the solar
neighborhood has produced enough $e^\pm$ in a not distant past, which
diffuse to Earth, leading to the observed excess\cite{Ioka10}.
Such a model can interpret both the total $e^\pm$ excess and the
positron fraction excess. Furthermore, depending model parameters,
both a sharp feature measured by ATIC and a smooth feature measured
by Fermi can be reproduced\cite{Ioka10}.

There are however two caveats related to this scenario.
First, it is not clear whether our galaxy is suitable to host
cosmological GRBs. Observations show that cosmological GRBs
typically reside in host galaxies that are different from
ours\cite{Fruchter06Nature,Savaglio09ApJ}. Some augthors argued
that Milky Way is too metal rich for bright GRBs\cite{Stanek06AcA},
while some recent observations indicate that at least some GRBs can
reside in host galaxies with metallicity as high as 0.3 solar
value\cite{DAvanzo10}. More data are needed to settle down the debate.
Second, even if bright GRBs can occur in
our galaxy, it is non-trivial to deposit nearly same amount of
$\gamma$-ray energy to TeV $e^\pm$ pairs that escape from the GRB.
One may envisage several pair production mechanisms in a GRB:
(i) A baryonic fireball may be pair-rich in the early stage. However,
most pairs would have annihilated when the fireball reaches the
$\gamma$-ray radiation radius.
(ii) In the prompt $\gamma$-ray emission region, TeV photons would
annihilate with hard X-ray photons to produce TeV pairs, but these
secondary pairs would quickly cool and re-radiate in the MeV
regime;
(iii) Prompt MeV photons may annihilate with the soft gamma-rays
back-scattered by the circum-burst medium to produce pairs.
These pairs have much lower energies. They may be reaccelerated
by the external shock, but the escaping fraction may not be high.
(iv) TeV photons emitted by the GRB (likely from external shock
since the compactness parameter is too high for internal prompt
emission) may interact with the ``background'' UV/optical photons
to produce TeV pairs. These pairs can most easily escape.
The question is how to get ``background'' UV/optical photons.
Prompt bright optical flash similar to that detected in the
naked-eye burst GRB 080319B\cite{Racusin08Nature} was invoked in
the model of Ref.\refcite{Ioka10}.
Such bursts are very rare, so that the chance of having a nearby GRB
capable of accounting for the observed $e^\pm$ excess is further
reduced.

\section{Scenario III: New Physics}
Although the electron/positron excess can be explained within the
astrophysical scenarios, another widely discussed scenario is
that the excess is the signature of dark matter.
The dark matter models have strong astrophysically motivations,
and can interpret the excess data under the assumption of some
(extreme) parameter regimes.
In this section, we critically review these models.
We begin with a brief review of the
observational evidence of dark matter, and
some widely-discussed dark matter candidates
(see also
Refs. \refcite{Jungman96PR,Bergstrom00RPP,Munoz04IJMPA,Bertone05PR,Hooper:2007qk,Olive08ASR,Steffen09EPJC,Hooper09Rev,arXiv:0907.1912,Bergstrom09NPJ,Ellwanger09,Feng09ARAA}
for extensive reviews on dark matter candidates).
We then go through various dark matter models (both annihilation
and decay models) proposed in the literature to interpret the
electron/positron excesses, and comment on the observational
constraints to the parameter spaces of these models. For a
detailed review on the annihilation models, see also
Ref.\refcite{He09IJPA}.

\subsection{Observational evidence for dark matter}
It was recognized as early as 1930s\cite{Zwicky:1933} that
some matter in the Universe is invisible. The non-baryonic nature
of dark matter was established since early 1970s\cite{Peacock}.
In the following we summarize
several compelling arguments for the existence
of non-baryonic dark matter.

\subsubsection{Galactic rotation curves}
\begin{figure}
\centerline{\psfig{file=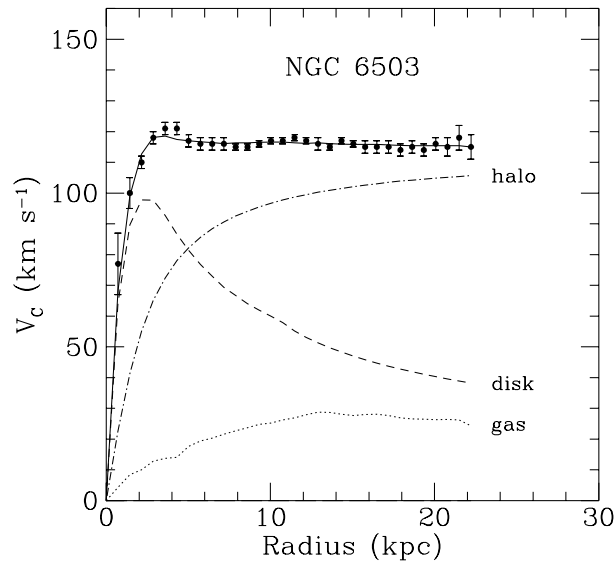,width=8cm}}
\vspace*{8pt}
\caption{Measured rotation curve of NGC6503 with best fit and the
  contributions from the dark matter halo, disk and gas (from Ref.153).
  }
\label{fig:rotationcurve}
\end{figure}

The most direct evidence for dark matter in the galactic scale comes
from the observations of the rotation curves of spiral galaxies (i.e.,
the graph of circular velocities $v_{\rm c}$ of stars and gas as a
function of their distance $r$ from the galactic center), which are
usually obtained by measuring Doppler shifts of
the 21 cm emission line of hydrogen gas.
The detected rotation curves usually exhibit a characteristic flat
behavior at a distance that is beyond the edge of the visible disks
(see Fig. \ref{fig:rotationcurve} for illustration). This implies
that the
mass $M_{<}(r)$ enclosed within a radius $r$
follows $M_{<}(r) \propto r$ for $r \gg r_{\rm disk}$, suggesting
the existence of an extended dark matter halo.
Measuring stellar velocity dispersion in elliptical galaxies
also leads to a similar conclusion.

\subsubsection{Galaxy Clusters}

Galaxy clusters are the largest gravitationally bound objects in the
Universe. Three independent methods all suggest that dark matter is
the main mass that bind clusters together.

First, Newton's gravity law suggests that
the velocity dispersion of
galaxies is approximately $v^2(r) \sim G M_{<}(r)/r$.  One
can infer the total mass $M_{<}(r)$ from the measurements of the velocity
dispersion and the size of a cluster (which can be determined if the
redshift and angular size of the cluster are measured).
In 1933, Zwicky\cite{Zwicky:1933} used this method to measure
the mass-to-light ratio of the Coma cluster to be 400 times of the
solar value.
This gave the first hints of dark matter in the modern sense.
The conclusion is confirmed by many follow-up measurements.

Second, according to the virial theorem, the gas temperature of the
clusters may be estimated as $k_{\rm B}T\approx 1.5~{\rm keV}~({{M_{<}{(r)}}
/ 10^{14}~M_\odot})({1~{\rm Mpc} / r})$, where $M_{<}(r)$ is
normalized to the baryonic mass derived from the star light $M_{b}
\sim 10^{14}~M_\odot$. However, the detected temperature is as high as
$\sim 10~{\rm keV}$, suggesting the existence of much more massive
dark matter, i.e., $M_{<}(r) \sim 6M_{b}$.

Finally, mass can be measured via the gravitational lensing effect
predicted by Einstein's general theory of relativity\cite{Blandford92ARAA}.
The mass of a lensing cluster can be inferred by the
deflection angle $\theta \approx ({G M_{\rm cl}/ d c^2})^{1/2}$
and the impact parameter $d$, both can be measured or inferred
if the redshift of the lensing cluster is known.
Usually the derived $M_{\rm cl}$ is much larger
than the observed baryonic mass $M_b$, again indicating the
existence of the large amount of invisible matter.

\begin{figure}[t]
  \includegraphics[scale=0.5]{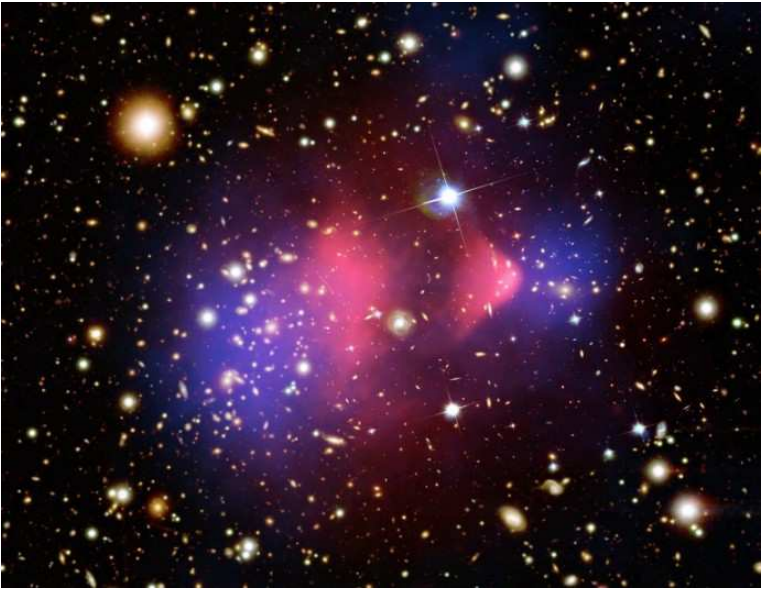}
  \caption{\label{fig:bulletcluster}
  The X-ray image by Chandra X-ray Observatory of the bullet cluster (red) over-imposed on the mass
distribution (blue) derived from gravitational lensing. It is clearly shown
that most of the mass does not trace the gas emission in the X-ray
band. This Figure is from http://www.spaceimages.com/hubucl1e06ph.html.}
\end{figure}

The most strong support to the dark matter picture came recently
from an X-ray observation of the bullet cluster
(Fig.\ref{fig:bulletcluster}). The X-ray image (red) shows the
distribution of baryonic mass (in the form of hot gas). The bulk of
mass distribution as inferred by the gravitational lensing technique
is denoted in blue, which does not trace the distribution of hot
gas. Such a configuration is difficult to interpret within other
competing models (e.g. the modified Newtonian dynamics), and
strongly supports the dark matter interpretation.

\subsubsection{Cosmological constraints on energy densities}

Advances in technology have usher in the new age of precision cosmology.
Wilkinson Microwave Anisotropy Probe (WMAP) can precisely measure
the primordial fluctuation in the temperature map of the cosmic
microwave background (CMB). By precisely locating the first acoustic
peak in the power density fluctuation, WMAP clearly measured the
curvature of the unvierse\cite{Spergel03}. Together with the
observational results of Type Ia supernova that suggest an
accelerating universe\cite{Riess98}, very precise values of the
parameters of the flat $\Lambda$CDM have been obtained.
According to the latest 7-year WMAP data\cite{Komatsu10},
the energy densities of various components of the $\Lambda$CDM
model are $\Omega_{\rm b} =0.0456\pm 0.0016$, $\Omega_{\rm dm} =
0.227\pm 0.014$, and $\Omega_\Lambda=0.728 \pm 0.015$.
An independent constraint on $\Omega_{\rm m}=\Omega_{\rm dm}+\Omega_{\rm b}$
from the baryon acoustic peak is consistent with the WMAP
result\cite{Eisenstein05}. All these strongly suggest that the
universe is dark-matter dominated, and that the bulk of dark matter
is not of the baryonic origin.
An independent constraint on the amount of barynic contribution
to dark matter was derived from the MAssive Compact Halo Objects
(MACHO) experiment, which showed that the micro-lensing detected
MACHOs can only account for a small fraction ($\sim 20\%$) of
the missing mass in the Milky Way and LMC halos\cite{Alcock00}.

\subsection{Dark matter properties and possible candidates}
The evidence for dark matter is compelling in all astrophysical
scales. To be a viable dark matter candidate, one needs to
satisfy some general requirements\cite{arXiv:0907.1912}: (1) Dark
matter must be \emph{dark}, in the sense that it must have no (or
extremely weak) interactions with photons. This is effectively
saying that the electric charge of dark matter particles must be
extremely smaller than the electron charge.
(2) Self-interactions of the dark matter should be rather weak.
This is manifested in the dark matter map of the Bullet Cluster
(Fig.\ref{fig:bulletcluster}). The collision between two galaxy
clusters cause strong shocks of baryonic gas between the two
clusters, while the two dark matter halos streamed through
each other without noticeable interaction.
(3) Interactions between dark matter and baryons must be very
weak. Otherwise baryons and dark matter would fall together
in the overdense region and form a baryon-DM disk in galaxies
rather than forming an extended DM halo.
(4) Dark matter cannot be mainly made up of standard model
particles, since most leptons and baryons are charged.  The only
potential candidate is neutrinos, but they are too ``hot'' and
too light to be trapped in DM halos.

Physicists are never short of ideas of inventing
non-baryonic dark matter candidates\cite{Bergstrom09NPJ,Feng09ARAA}.
Almost all current models of dark matter use the standard concept of quantum
field theory to describe the properties of elementary particle candidates
(for exceptions, see for instance Refs.\refcite{unparticles,qballs}). This
means that the dark matter particles can be characterized by their
mass and spin. In the following we summarize the properties of several
widely discussed dark matter candidates. As shown in Table~\ref{tab:DM-1},
the mass of the proposed candidates spans a wide range.

The most studied dark matter candidates are weakly-interacting massive
particles (WIMPs). The main motivation for introducing WIMPs is the gauge
hierarchy problem, namely, the observational constraint of the mass of
the spin 0 Higgs Boson (a hypothetical massive particle in the
Standard Model (SM) in particle physics), i.e. $114.4 ~{\rm GeV}<
m_{\rm Higgs} < 186 {\rm GeV}$\cite{LEP03PLB,LEP}, is much lighter
than the Planck mass $M_{\rm Pl}=1.2\times 10^{19}
{\rm GeV}$. This suggests that SM is not the complete model, and there
should be new physics in the energy range of the weak interaction
scale $m_{\rm weak} \sim 10~{\rm GeV - TeV}$. WIMPs are the predicted
particles of the new physics, which naturally carries a mass with the
similar energy scale. Very interestingly, this energy scale is broadly
consistent with the energy scale of the PAMELA/ATIC/HESS/Fermi $e^\pm$
excess. This was the main reason that the $e^\pm$ excess stimulated
the great excitement from the particle physics community. Three types
of WIMPs, i.e. neutralinos and other supersymmetric particles,
Kaluza-Klein particles in universal extra
dimensions, and inert Higgs doublets, are discussed below. Some
candidates, such as axions and sterile neutrinos, have too low an
energy to be relevant to the $e^{\pm}$ excess data. Since
they are physically well motivated, we
nonetheless include them for the sake of completeness.

\begin{table}[h]
\tbl{Properties of some Dark Matter Candidates (adapted from Refs.147, 149).
}
{\begin{tabular}{@{}lll@{}} \toprule
Type&Particle Spin&Approximate Mass Scale \\ \colrule
Neutralino&1/2&10 GeV - 10 TeV\\
Kaluza-Klein UED&1&TeV\\
Inert Higgs Doublet&0& $\sim $50 GeV or $\sim$650 GeV\\
Axion& 0&$\mu$eV-meV\\
Sterile Neutrino&1/2&~keV\\
\colrule
\end{tabular}}
\label{tab:DM-1}
\end{table}

\subsubsection{Neutralinos and other supersymmetric particles}\label{sec:mssm}

The most widely discussed WIMPs are neutralinos, one type of the
supersymmetric (SYSY) particles.  Supersymmetry (SUSY) is an
ingredient in many superstring theories trying to unite all the
fundamental forces of nature, including gravity\cite{wess}. The gauge
hierarchy problem is most elegantly solved by SUSY. In SUSY extensions
of the standard model, every standard model particle has a new, as yet
undiscovered partner particle, which has the same quantum numbers and
gauge interactions, but differs in spin by $1/2$.
From the standpoint
of dark matter, the lightest supersymmetric particle is naturally stable in models
that conserve $R$-parity, which is defined as
\begin{equation}
R=\left(-1\right)^{3{\cal B}+L+2S},
\end{equation}
where ${\cal B}$ is the baryon number, $L$ the lepton number and $S$
the spin of the particle\cite{Jungman96PR,Bertone05PR,Bergstrom09NPJ}.
The ordinary particles
and their superpartners have $R=+1$ and $-1$ respectively, which means
that supersymmetric particles can only be created or annihilated in
pairs in reactions of ordinary particles, or a single supersymmetric
particle can only decay into final states containing an odd number of
supersymmetric particles (plus some Standard Model particles).  The idea
that supersymmetric particles could be good dark matter candidates
became attractive when it was realized that breaking of SUSY could be
related to the electroweak scale, and that the supersymmetric partner
of the photon (the photino) would couple to fermions with electroweak
strength\cite{fayet}.  Below we list some supersymmetric dark matter
candidates\cite{ellis0,Bertone05PR} with brief discussion.

\begin{itemize}
\item {\it Neutralinos} in models of R-parity conserving supersymmetry
are by far the most widely studied dark matter candidates. Since the
superpartners of the $Z$ boson (zino), the photon (photino) and the
neutral higgs (higgsino) have the same quantum numbers, they can mix
to form four eigenstates of the mass operator called
``neutralinos''\cite{haberkane}.
In many models the lightest of the four neutralinos
$\chi$ turns out to be the lightest supersymmetric particle (LSP),
which is, therefore, stable and can only be destroyed via pair
annihilation (Neutralinos have spin $1/2$ and are their own
antiparticles; that is, they are Majorana fermions). $\chi$ can be
produced thermally in the hot early universe and leave approximately
the right relic abundance to account for the observed $\Omega_{\rm
dm}h^2\sim 0.1$, where $h\sim 0.71$ is the Hubble's constant in units of $100~{\rm km~s^{-1}~Mpc^{-1}}$. These facts make $\chi$ with a mass of roughly
10-10000 GeV, an excellent cold dark matter candidate.  They may be
detected directly through scattering in detectors, or indirectly
through the decay products that result when neutralinos annihilate in
pairs\cite{Jungman96PR,Bertone05PR}.

\item{\it Sneutrinos} are the superpartners of the
Standard Model neutrinos in the supersymmetric models. They have a
spin $0$, and have long been considered as dark matter
candidates\cite{ellis0}. If their mass is
in the range of $0.55-2.3$ TeV, the sneutrinos can have a
cosmologically interesting relic density\cite{Falk94PLB}. From the
direct detection searches sneutrinos have been excluded as the
major component of the dark matter\cite{Falk94PLB}. However, in the
supersymmetric models with the inclusion of right-handed fields,
sneutrinos are still compatible with the current direct detection
sensitivities and could be a viable dark matter
candidate\cite{Arina07JHEP}.

\item{\it Gravitinos}
are the spin $3/2$ superpartners of gravitons in the supersymmetric
models, and could be the lightest, stable supersymmetric particle.
Although strongly theoretically motivated, gravitinos as dark matter
would be very difficult to observe due to their extremely weak interactions
with ordinary matter\cite{Feng03PRL}. Long lived gravitinos may
impose problems to cosmology. For example, their presence may destroy
the abundances of primordial light elements\cite{Moroi93PLB,Bertone05PR}.
These problems can be
circumvented in some scenarios\cite{Buchmuller07JHEP,Steffen09EPJC}.
For instance, in the case of a broken R-parity, the constraints from
primordial nucleosynthesis are naturally satisfied and decaying
gravitinos would lead to characteristic signatures in high energy cosmic
rays\cite{Buchmuller09JCAP}.
\end{itemize}

\subsubsection{Kaluza-Klein Particles in Universal Extra Dimensions}
Our world appears to consist of $3+1$ dimensions (three spatial
dimension and one time dimension). It is however possible that there
exist other $\bar{\delta}$ spatial dimensions, which
appear at much higher energy scales, as suggested by Kaluza and Klein
in the 1920s. In the extra-dimension models, the $(3 + 1)$-dimension
space-time one experiences is a structure called a ``brane'', which is
embedded in a $(3 +\bar{\delta} + 1)$ space-time called the ``bulk''.
Upon compactification of the extra dimensions, all the fields
propagating in the bulk have their momenta quantized in units of
$p^{2} \sim 1/\Re^{2}$, where $\Re$ denotes the size of the extra
dimensions\cite{Bertone05PR}. As a result, for each bulk field a set
of Kaluza-Klein (KK) states (i.e. Fourier expanded modes) appear.
In the 3+1 dimensional world, these KK states appear as a series of
states with masses $m_{n} = n/\Re$, where $n$ labels the mode
number. Each of these new states contains the same quantum numbers,
such as charge, color, etc\cite{Bertone05PR}. Usually the standard
model fields are assumed to be confined on the brane and only
gravity is allowed to propagate in the bulk. However it might be
possible for all the standard model particles
freely propagate in the extra dimensions\cite{Cheng02PRD1}.
Such a scenario is the so-called universal extra dimensions.
The simplest universal extra dimension models preserve a discrete
parity known as the KK-parity, implying that the lightest KK particle
(LKP) is stable and can be an interesting, possible dark matter
candidate\cite{Bertone05PR,Hooper:2007qk}. The LKP is typically the
level 1 partner of the hypercharge gauge bosons, denoted as
$B^{(1)}$~\cite{Cheng02PRD1}.  The WMAP limit of $\Omega_{\rm dm}
h^2 = 0.1131 \pm 0.0034 $\cite{Komatsu10} corresponds to a mass of the
dark matter candidate $B^{(1)}$ between roughly 0.5 to 1 TeV,
depending on the exact form of the mass spectrum and the resulting
co-annihilation channels\cite{Servant03NPB,Bergstrom09NPJ}. Current
collider measurements give a constraint of $\Re^{-1} \geq 0.3\ {\rm
TeV}$, whereas LHC may probe the KK mass parameter space up to
$\Re^{-1}\sim 1.5$ TeV\cite{Hooper:2007qk}.

\subsubsection{Inert Higgs}
Inert Higgs are a type of WIMPs that are introduced in
one of the most minimal extensions of the standard model.
Back in 1978, it was noted that
a model with two Higgs doublets containing a discrete symmetry
could contain a state, the lightest neutral scalar or pseuodoscalar
boson, which is stable\cite{ma,ma-2}. It is the subject of the
renewed interest recently since it could be used to solve the
naturalness problem in the standard model\cite{rychkov}.
The possibility of one of the lighter neutral states
in the enlarged Higgs sector to be the dark matter was also pointed
out, and soon the basic properties of this ``inert'' Higgs candidate
for dark matter were investigated\cite{honorezlopez}. It turns out
that this model contains a dark matter candidate, an ``inert'' particle
that couples gauge bosons only and is stable due to its discrete
symmetry. Taking into account the relic density constraints from WMAP
and various theoretical and experimental constraints, the latest
constraint on the mass of the Inert Higgs is $\sim 30-80$ GeV (middle
mass range) or $\sim 500-800$ GeV (high mass range)\cite{Dolle09PRD}.
Candidates in the middle mass range may annihilate through $Z$ or
through Higgs. Some solutions in the middle mass range may also
have a large annihilation cross section, through loop corrections,
into a pair of gamma rays. This is relevant for indirect detections
of dark matter through annihilation at the Galactic Center, and
such a gamma ray line would be easily observed by the Fermi
satellite\cite{Gustafsson07PRL}. For the inert Higgs dark matter
candidates in the high mass range, annihilation into $W^{\pm}$ pairs
is allowed with a large cross section\cite{honorezlopez}.

\subsubsection{Axions}\label{subs:axions}
Axions were introduced in an attempt to solve the strong $CP$ problem
in particle physics\cite{PQ}. They are still one of the leading
dark-matter candidates.
The axion mass is defined by
\begin{equation}
m_{\rm a}={\sqrt{m_{\rm u}m_{\rm d}}\over m_{\rm u}+m_{\rm d}}m_\pi f_\pi
{1\over f_{\rm a}}\approx 0.62\ {\rm eV}\left({10^7\ {\rm GeV}\over
f_{\rm a}}\right),
\end{equation}
where $m_{\rm u}\simeq 4$ MeV, $m_{\rm d}\approx 8$ MeV and
$m_\pi\approx 135$ MeV are masses of the up quark, down quark and pion,
and $f_\pi\approx 93$ MeV is the pion decay constant.  The axion
phenomenology is therefore determined by, up to a numerical factor,
by one number only, i.e. the energy scale $f_{\rm a}$ of symmetry
breaking. A variety of
astrophysical observations and laboratory experiments constrain
$f_{\rm a}$ to be between $10^{9}~{\rm GeV}$ and $10^{12}~{\rm GeV}$,
suggesting an axion mass $m_{\rm a}\sim 10^{-5}-10^{-2}$ eV. Smaller
masses would lead to an unacceptably large cosmological abundance,
while larger masses are ruled out by combinating of constraints from
supernova 1987A, globular clusters, laboratory experiments, and the
search for two-photon decays of relic
axions\cite{Kim09RMP,Bergstrom09NPJ}. The lifetime of axion is
$\tau({\rm axion} \rightarrow \gamma \gamma)\approx {9\times 10^{23}~{\rm s}
\over g_\gamma^{2}}({{\rm1~ eV}\over m_{\rm a}})^{5}$, where $g_\gamma$ is
a model-dependent parameter and $g_\gamma^{2}$ is expected not to
deviate from 1 significantly. In general, its very small mass makes
axion not possible to interpret the 100 GeV $e^\pm$ excess discussed
in this review.

Within the SUSY theories, axions have spin $1/2$ superpartners,
which are called axinos. Models
combining SUSY and the Peccei-Quinn\cite{PQ} solution to the strong CP
problem necessarily contain this particle. Depending on the model and
the SUSY breaking scheme, the axino mass can range from eV to
GeV. Although widely believed to be only capable of acting as a warm, or
hot dark matter candidate\cite{Goto92PLB,Bonometto94PRD}, it was
recently shown that for quite low reheating temperatures, cold
axino dark matter may be possible\cite{Covi99PRL,Chun00PRD}.
Interesting astrophysical signals are expected from axinos
if the R-parity is broken\cite{Covi09NJP}.

\subsubsection{Sterile Neutrinos}
The Standard (active) neutrinos were considered as a favored
particle dark matter candidate in late 1970's, when
the calculations of the massive neutrino relic density were
first carried out\cite{hut,leeweinberg,vysotsky,gunn}.
To account for the dark matter
of a dwarf galaxy of velocity dispersion $\sigma_{\rm d}$ typically of
order $100~{\rm km~s^{-1}}$ and core radius $r_{c}$ typically $1~{\rm
kpc}$, the neutrinos have to be ``massive" \cite{tg,arXiv:0907.1912}
\begin{equation}
m_{\nu}\geq 120\ {\rm eV}\left({100\ {\rm km/s}\over\sigma_{\rm
d}}\right)^{{1\over 4}}\left({1\ {\rm kpc}\over
r_{c}}\right),\label{eq:gt}
\end{equation}
which is much larger than the upper limit $m_{\nu}<2.3$ eV
derived from the various experimental results (e.g. the
tritium $\beta-$decay experiments at Mainz\cite{Kraus05}.
Therefore the standard neutrinos are too light to
be the dominant component of dark matter, although among the
proposed candidates for non-baryonic dark matter standard
neutrinos have the ``undisputed virtue of being known to
exist''\cite{Bergstrom00RPP}.

Sterile neutrinos, unlike standard (active) neutrinos, do
not interact through standard weak
interactions\cite{sterile,Shi98PRL}. Instead, they communicate with
the rest of the neutrino sector through fermion
mixing\cite{abazajian,shaposh}.  The existence
of sterile neutrinos (right-handed or gauge singlet) is one of the
most attractive explanations of the observed flavor oscillations of
active neutrinos. Sterile neutrinos were proposed as dark matter
candidates in 1994\cite{sterile}. Stringent cosmological and
astrophysical constraints on sterile neutrinos come from the analysis
of their cosmological abundance and the study of their decay
products\cite{Abazajian01PRD}.  A very recent analysis finds out that
sterile neutrino DM with mass $\geq 2$ keV is consistent with all
existing constraints\cite{Boyarsky09PRL}. On the other hand, this
mass is too low to account for the 100 GeV $e^\pm$ excess feature
discussed in this review.

\subsection{Dark matter models for the $e^\pm$ excess}
In many dark matter models, the dark matter particles are their own
anti-particles (Majorana fermions), and can annihilate with each
other.  There are several possible DM annihilation scenarios:
(i) DM particles annihilate to standard model particle pairs, such
as gauge bosons, quarks, and lepton
pairs\cite{Jungman96PR,Bertone05PR,Cirelli09NPB,Barger09PLB,Cholis09PRD,Cirelli09NPB-1};
(ii) DM particles annihilate via virtual internal bremsstrahlung
processes to produce $e^{+}e^{-}\gamma$\cite{Bergstrom08PRD};
(iii) DM particles annihilate to new mediating particle
pairs, which then decay to $e^{+}e^{-}$ and other standard model
particles\cite{arkani,Pospelov09PLB,Cholis09JCAP}.
According to these scenarios, dark matter annihilation would lead
to electron/positron pairs, and may account for the PAMELA/ATIC/HESS/Fermi
$e^\pm$ excess (see e.g.,
Refs.\refcite{Dicus77PRL,Tylka89PRL,Turner90PRD,Jungman96PR,Baltz99PRD,Bertone05PR,He09IJPA,arkani,leptophilic,lep1,lep2,lep3,bi-he},
\refcite{Bergstrom08PRD,susy1,susy2,susy3,susy4,susy5,susy6,susy7,susy8,susy9,susy10,susy11,susy12,susy13,susy14,susy15,susy16,susy17},
\refcite{nmssmhooper,nmssmhooper1,nmssmlykken}, \refcite{kk,kk1,kk2},
\refcite{light,light1,Cholis09PRD,light3,light4,light5,light6,light7,Cholis09JCAP,light9,light10,light11},
\refcite{light12,light13,light14,light15,light16,light17,light18,light19,light20,light22,light23},
\refcite{annihilation,annihilation1,annihilation2,annihilation3,annihilation4,annihilation5},
\refcite{annihilation6,annihilation7,annihilation8,annihilation9,annihilation10,annihilation11,annihilation12,annihilation13,annihilation14,annihilation15,annihilation16}).
According to these scenarios, the resulting lepton spectra would have
a cut off energy at the DM mass $m_{\rm dm}$.  Hence, the cutoff
energy seen in Fermi/HESS (or ATIC) data would suggest
$m_{\rm dm} \sim 1.5$ (or 0.6) TeV.

The dark matter self-annihilation scenarios, which have been
extensively discussed since 1970s, are not the only
possibilities to produce signatures for dark matter indirect
detections.  Strictly speaking, the viability of a particle as
a dark matter candidate does not require its absolute stability,
but merely require that the dark matter lifetime
$\tau_{\rm dec}$ is much longer than the age of the Universe (i.e.,
$\tau_{\rm dec}\gg 10^{18}$ s). Hence, if dark matter decays
at a sufficiently high rate, the decay products might be
detectable.  In particular, if the DM decays predominately into
leptons in a time scale $\tau_{\rm dec} \sim 10^{26}$ s (which is
long enough not to cause other cosmological problems), the detected
100 GeV $e^\pm$ excess may be also reproduced (see e.g.,
Refs.\refcite{Buchmuller07JHEP,Ibarra08PRL,Yin08PRD,decay,decay1,decay2,decay3,decay4,decay5,decay6,decay7,decay8,decay9,decay10,decay11,decay12},
\refcite{decay13,decay14,decay15,decay16,decay17,decay18,decay19,decay20,decay21,decay22,decay23,decay24,decay25,decay26,decay27,decay28},
\refcite{decay29,decay30,decay31,decay32,decay33,decay34}).
In these DM decay scenarios, the resulting $e^\pm$ spectra would
have a cutoff at an energy $\sim m_{\rm dm}/2$.  Hence, the cutoff
energy seen in Fermi/HESS (or ATIC) data would suggest
$m_{\rm dm}\sim 3$ (or 1.2) TeV.

In both the annihilation and decay scenarios, the DM required in
the modeling belongs to the WIMP category. This is encouraging since
weak-scale particles make an excellent dark matter candidate,
having the merit of the so-called WIMP miracle:
In the simplest type of models
of thermally produced dark matter, an average of the annihilation rate
at the time of chemical decoupling can be estimated as $<\sigma v>
\approx 3\times 10^{-26}(0.1/\Omega_{\rm dm}h^{2})~{\rm cm}^{3}~{\rm
s}^{-1}$, where $<\sigma v>$ is the thermal average of the
total annihilation cross section $\sigma$ multiplied by velocity $v$.
Taking the relative velocity of colliding DM particles at the
decoupling time $v\sim 10^{10}~{\rm cm~s^{-1}}$, one has $\sigma \sim
10^{-8}~{\rm GeV}^{-2} \sim \alpha^{2}/m_{\rm dm}^{2}$, or
$m_{\rm dm} \sim 100[\alpha/(1/137)]~{\rm GeV}$. This mass is
in the energy range of weak interaction energy scale, and hence
the particle is likely a WIMP.

In the following, we review the annihilation and decay models
in turn.

\subsubsection{Dark matter annihilation models}
In the dark matter annihilation scenarios, the source term of the
electrons/positrons is given by
\begin{equation}
q(\vec{r},p)={1\over 2}\left({\rho_{\rm dm}(\vec{r})\over m_{\rm
dm}}\right)^{2}<\sigma v>{dN_{\rm e^\pm} \over d{\cal E}},
\label{eq:Q_annihilation}
\end{equation}
where ${dN_{\rm e^\pm} \over d{\cal E}}$ is the energy
spectrum of the electrons/positrons produced by DM annihilation,
$\rho_{\rm dm}(\vec{r})$ is the dark matter density distribution in the
annihilating region, and $<\sigma v>\sim 3\times 10^{-26}~{\rm
cm^{3}~s^{-1}}$ based on the WIMPs miracle argument discussed above
(i.e. the usual DM annihilation rate producing
the $e^\pm$ excess signal is also related to the annihilation rate
producing the cosmological relic DM density). With a given
$q(\vec{r},p)$, one can solve the propagation equation (i.e.,
Eq.[\ref{eq:propagation}]) and get the final electron/positron
spectra detected on Earth. For such a purpose, the density
distribution of the dark matter halo is needed. Various models
of DM halo density distribution have been proposed. Some popular
ones can be parameterized as
\begin{equation}
\rho_{\rm dm}(\vec{r})=\rho_{\rm dm,\odot}\left({r_\odot \over
r}\right)^{\Upsilon}\left[{1+(r_\odot/r_{\rm s})^{\omega}\over 1+(r/r_{\rm
s})^{\omega}}\right]^{(\zeta-\Upsilon)/\omega},
\end{equation}
where $\rho_{\rm dm,\odot} \sim 0.3~{\rm GeV~cm^{-3}}$ is the DM
density in the
vicinity of the sun, $r_\odot \sim 8.5$ kpc is the solar
distance from the Galactic center, and the profile parameters
$(\omega,~\zeta,~\Upsilon,~r_{\rm s})$ are $(2,~2,~0,~5~{\rm kpc})$,
$(1,~3,~1,~20~{\rm kpc})$, $(1,~3,~1.5,~30~{\rm kpc})$ for the Core
Isothermal model\cite{Bahcall80ApJ}, the Navarro, Frenk and White
(NFW) model\cite{NFW} and the Moore model\cite{Moore99MN},
respectively. It is well known that only the TeV CR $e^\pm$
produced within a distance $\sim 1$ kpc can reach us, i.e., $r <<
r_{\odot}$, hence the above 3 popular (but different)
dark matter density distribution
models give rise to rather similar CR $e^\pm$ signals.
The photon signals from these three different models, such as the
GeV-TeV gamma-rays generated from inverse Compton scattering off the
CMB photons by the TeV $e^\pm$'s or the synchrotron radiation of
the TeV $e^\pm$'s, can however reach us from a much larger
distance range including the core of the DM halo. These signals
therefore can be very different for different models.
If DM annihilation also produces hadronic pairs,
the spectrum of the antiproton CRs can be also very different
for different models,
since antiprotons are more rigid and much less radiative so that
they can reach from much larger distances. The non-detections of a
prominent electromagnetic spectral signals in the radio (synchrotron
origin) and GeV-TeV (inverse Compton origin) bands, as well as the
non-detection of a prominent
antiproton cosmic ray signal, therefore impose stringent constraints on
the DM density profile and/or the DM annihilation properties.

Since the first release of the PAMELA data, many candidate WIMP DM
models have been studied.
The spectral shape of
the observed $e^\pm$ excess is not difficult to reproduce if the
annihilations proceed mostly to leptons rather than to quarks, gauge
bosons, and so on. For example, two-body annihilations directly into
$e^\pm$ pairs produce a hard spectrum, which can be adjusted to be
consistent with the ATIC data. If the two-body final states are
$\mu^\pm$ or $\tau^\pm$ pairs, their subsequent decays into $e^\pm$
can produce a soft enough spectrum consistent with the Fermi and
HESS data.

However, in all these models, the thermal average
cross section multiplied by velocity $<\sigma v>=3\times 10^{-26}~{\rm
cm^{3}~s^{-1}}$ is found to be too small to account for the flux
level detected by PAMELA/ATIC/Fermi/HESS. An unnaturally large
``enhancement/boost factor'', ${\rm E_{\rm F}}\sim 10^{2}-10^{3}$, is
needed. Quite a few mechanisms have been proposed to introduce
such a ${\rm E_{\rm F}}$ factor. (1) If DM halos have substructures,
the local density can be enhanced so that the annihilation optical
depth can be increased\cite{Lavalle08AA}. Detailed N-body simulations,
on the other hand, show that the boost factor introduced by DM
substructures is generally less than ${\rm E_{\rm F}} \sim 10$;
(2) It is possible that DM in the early universe was not in
equilibrium. In the scenarios of non-thermal production of DM, the
interaction rate responsible for the $e^{\pm}$ excesses are not
directly related to the relic DM density. One may simply assume
that the interaction rate is that required from the $e^{\pm}$
excess data, and no boost factor is needed\cite{Dutta09PRD};
(3) More physically, there might exist new mediating particles
during the DM annihilations. Examples include
the Sommerfeld mechanism\cite{sommerfeld,arkani} and the Breit-Wigner
mechanism\cite{Lattanzi09PRD}.
The Sommerfeld mechanism\cite{sommerfeld} is a non-relativistic
quantum mechanical effect. A large boost factor can be produced
if DM interacts with a new light particle. Without such an
interaction, annihilation of DM is a short distance effect.
If DM interacts with a nearly massless (light) particle, then
there is an additional long-range force (with Coulomb-like potential
$V(r)=-b/r$), which would distort the
DM wave-function before annihilation happens. This would introduce
a modification factor to the annihilation cross section, so that
$\sigma=\sigma^0 S$. The factor $S$ is the Sommerfeld factor.
By solving the non-relativistic Schr\"odinger equation, one can
derive $E_{\rm F}=S$.
Assuming a dark matter particle $\chi$ coupling to a
mediator $\phi$ with coupling strength $\lambda$, for S-wave
annihilation in the nonrelativistic limit, one gets ${\rm E_{\rm
F}}=(\pi b/ {\rm v})/(1-e^{-\pi b/{\rm v}})$ for $m_{\phi}=0$ and
${\rm E_{\rm F}} \leq \pi b m_{\rm dm}/m_{\phi}$ for $m_{\phi}>0$,
where
${\rm v}$ is the velocity
(in units of $c$) of each DM particle in the center-of-mass
frame\cite{arkani}. ${\rm E_{\rm F}} \sim 10^{3}$ is then achievable if
${\rm v} \sim 10^{-3} b \pi$ for $m_{\phi}=0$ or $m_{\rm dm}/m_{\phi}
\sim 10^{3}/b \pi$ for $m_{\phi}>0$.
The Breit-Wigner mechanism\cite{Lattanzi09PRD} applies to the case of
a heavy mediating particle. For most masses of the mediating particle,
the cross section is not enhanced. However, if the annihilation of
DM is through the S-channel and if the new particle has a mass about
twice of the DM mass, the cross section and annihilation rate can be
greatly enhanced near a resonance point. This introduces a large
enough $E_{\rm F}$ to interpret the data.

Another challenge the DM model faces in the $e^{\pm}$ excess
modeling is the non-detection of an excess in the antiproton
spectrum\cite{pamela-p}. Contrary to the data, most DM annihilation
models invoke a large hardronic annihilation fraction.

The requirement of having excesses only in $e^\pm$ but not in
anti-proton spectrum, along with the large boost factor needed
as discussed above,
pose great constraints and
eliminate many candidate DM annihilation models.  For
example, the most popular DM candidate, the lightest supersymmetric
particle (LSP) in the minimal
supersymmetric standard model (MSSM) can give rise to the $e^\pm$
excesses\cite{Bergstrom08PRD}.  However, as a linear combination of
photino, zino and higgsino, LSP usually has a large hadronic
annihilation fraction, which is in conflict with the PAMELA
anti-proton data. This model also suffers the problem of realizing
a large boost factor.  For the KK dark matter particles, direct
annihilation to $e^+e^-$
is possible. In fact, the lightest KK
particles (LKP) was recognized \cite{Cheng02PRD1} as a potentially
important source of positrons when the $e^\pm$ excess was found by
HEAT earlier. In the ATIC discovery paper\cite{atic}, the KK models
was also introduced as an example to account for the observed
$e^{+}+e^{-}$ spectrum peak around 600 GeV.
These models again require a significant fraction of annihilation in
hadronic modes ($\sim 35\%$), which is at odds with the PAMELA
antiproton data (cf. Ref.\refcite{kk1}). The required large boost
factor $E_{\rm F}$ is attributed to halo sub-structures, which was
found difficult to achieve.  In the inert Higgs model, for a candidate
with $m_{\rm dm} \sim 70$ GeV, the DM annihilation contribution to
$e^{+}$ and anti-proton fluxes is much smaller than the expected
backgrounds. Even if a boost factor is invoked to enhance the signals,
the particle candidate is unable to explain the observed $e^{+}$ and
anti-proton data. Only a high mass candidate, with $m_{\rm dm}\sim
10$ TeV, would be possible to fit the PAMELA excess provided
a large enhancement, either in the local DM density or
through the Sommerfeld effect\cite{Nezri09JCAP}. Such a heavy
particle, however, is in contradiction with the constraint from WMAP and
various theoretical and experimental constraints\cite{Dolle09PRD}.

\begin{figure}[t]
\includegraphics[width=0.48\columnwidth]{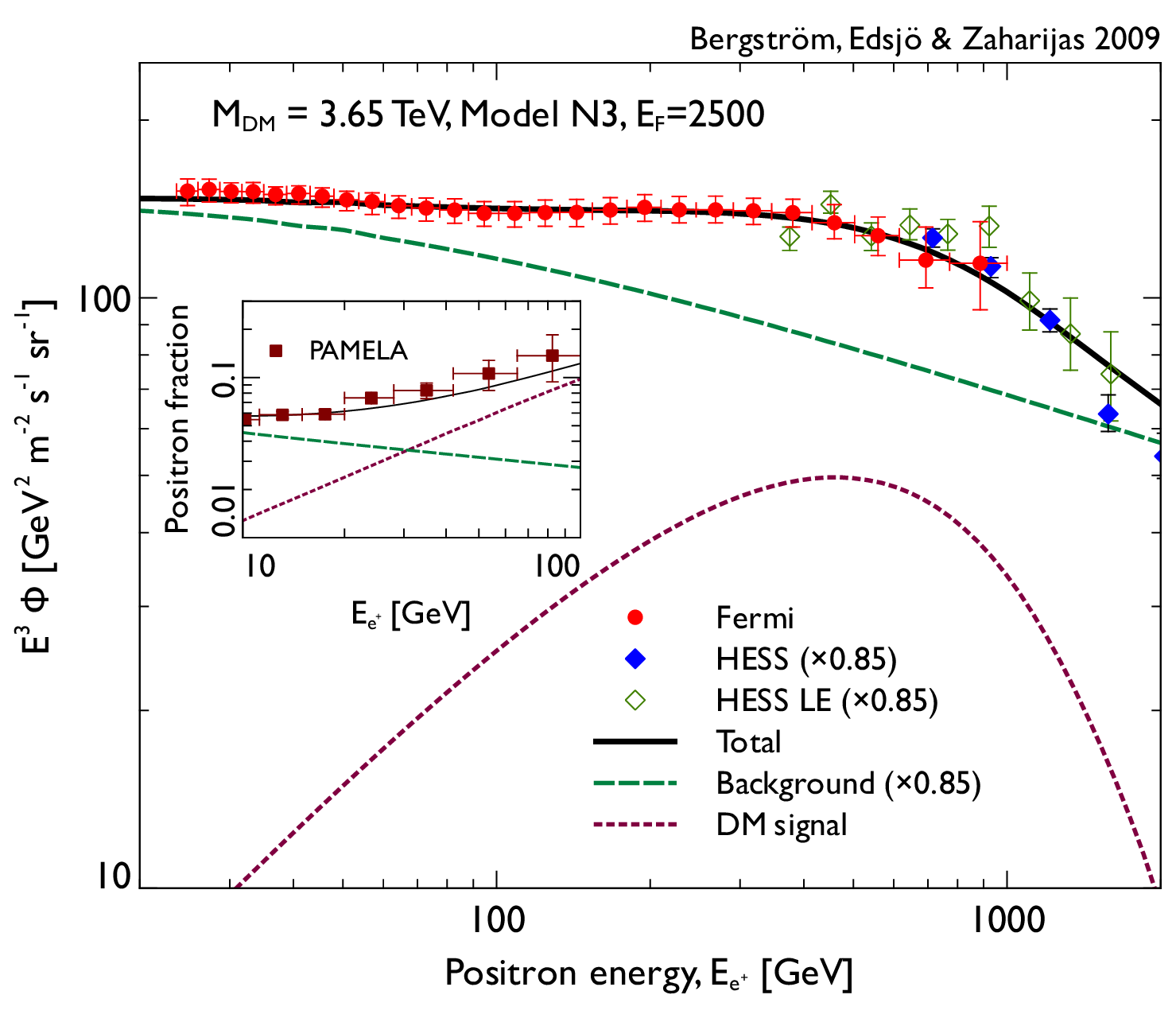}
\includegraphics[width=0.48\columnwidth]{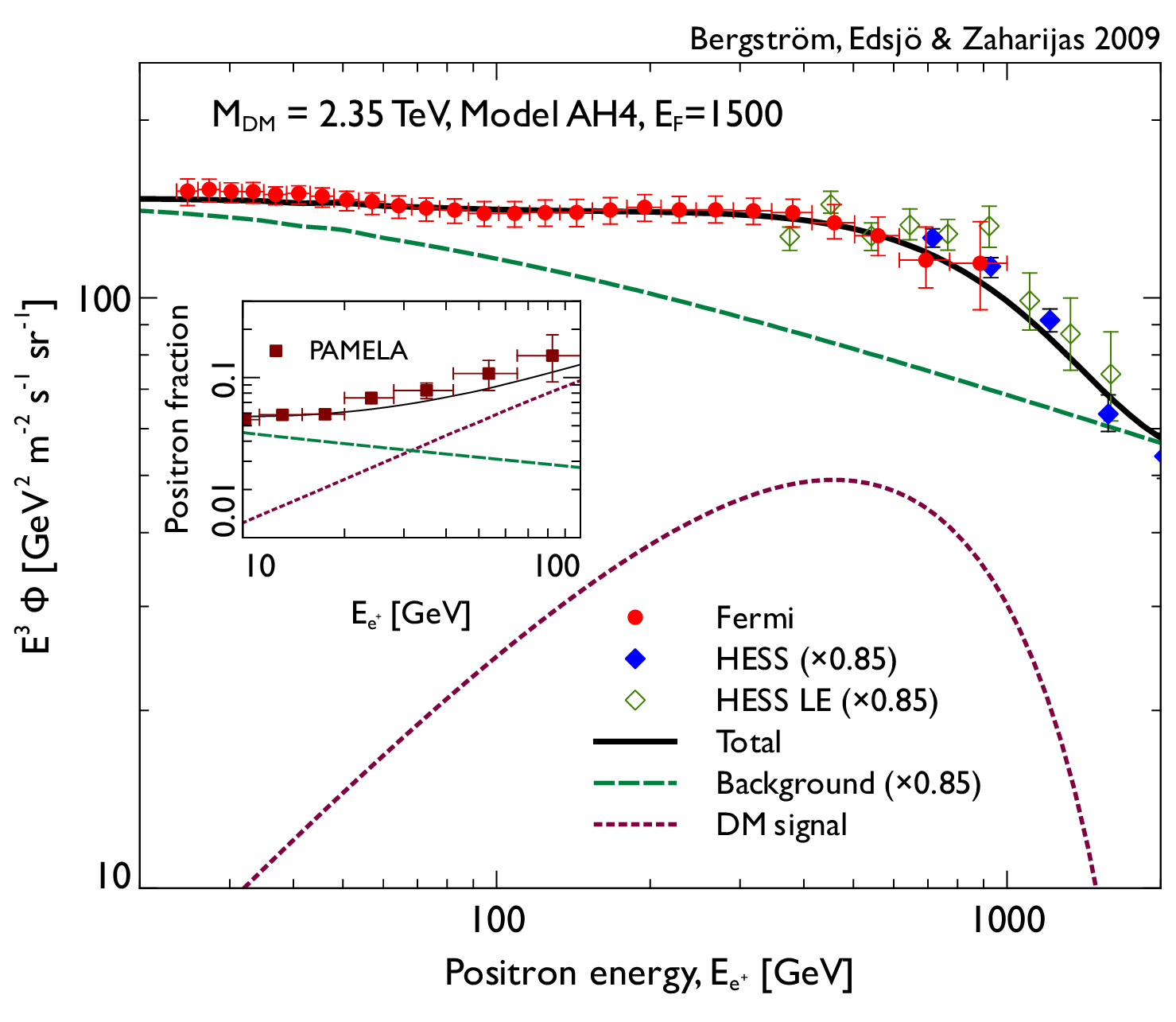}
\caption{
\label{fig:annihilation1}
Example spectra of good fit DM annihilation models found in
Ref.324.
The predicted $e^+ + e^-$ signal and background are compared against
the Fermi and HESS data. The HESS data and the background
model have been rescaled with a factor 0.85. In the inset, the positron
fraction as measured by PAMELA is plotted against the predicted
signal for the same model.
}
\end{figure}

Recently two types of models have been constructed to account for the
detection of the excess in the electron/positron channel but not in the
antiproton channel (see Ref.\refcite{He09IJPA} for a recent review):

(1) The kinematically limited light particle decay models (e.g.,
Refs.\refcite{arkani,light,light1,Cholis09PRD,light3,light4,light5,light6,light7,Cholis09JCAP,light9,light10,light11},
\refcite{light12,light13,light14,light15,light16,light17,light18,light19,light20,light22,light23});
These models require the existence of a light
particle with a mass less than the sum of proton and anti-proton
masses. Such a low mass makes hadronic modes kinematically
inaccessible and forces the particle decay predominantly into
leptons. To accommodate the very high annihilation rate required to
reproduce the current $e^{\pm}$ excess, the light particle decay
model makes use of its light force-carrier to generate the Sommerfeld
enhancement\cite{arkani}. For instance, in the MSSM extended by a
singlet chiral superfield, the singlino-like neutralino dark matter
annihilates to light singlet-like Higgs bosons, which proceed to decay
to either electron-positron or muon-antimuon pairs. For the
singlino-like neutralino dark matter with a mass $\sim 2$ TeV, PAMELA
and Fermi data may be interpreted\cite{nmssmhooper,nmssmhooper1}.
Phenomenologically, the light particle decay can have multiple
channels. For example, for the case of the mediating light particle
being a light scalar $\phi$\cite{arkani,Cholis09JCAP}, for annihilation
$\chi\chi \rightarrow \phi \phi$, one would consider four $\phi$
decay channels: 1) $\phi \rightarrow e^{+}e^{-}$, 2)
$\phi \rightarrow \mu^{+}\mu^{-}$, 3) a mixture of 1:1 (or 2:1)
between electrons and muons, and lastly 4) $\phi \rightarrow
\pi^{+}\pi^{-}$\cite{Cholis09JCAP}. Detailed modeling suggests that
the models annihilating predominantly to $\mu$ are somewhat
favored\cite{Bergstrom09PRL}. For the case of mediating light
particle is a scalar $s$ and a pseudoscalar $a$, i.e. $\chi\chi
\rightarrow s a$, one has $a$ also preferably decay to muons, i.e.
$a \rightarrow \mu^{+}\mu^{-}$, and muons further decay to $e^{\pm}$
detected on the Earth.  Some spectrum modeling
examples\cite{Bergstrom09PRL}
can be found in Fig.\ref{fig:annihilation1}, where the model
${\rm N_{3}}$ is one of the Nomura-Thaler-type models\cite{light3}
and is characterized by $m_{a}=0.8$ GeV and $m_{s}=20$ GeV. The model
${\rm AH_{4}}$ is one of the Arkani-Hamed et al.$-$type
models\cite{arkani}, in which the scalar dark matter particles
annihilate into $\mu^{+}\mu^{-}$.

(2) The leptophilic DM models\cite{leptophilic,lep1,lep2,lep3,bi-he}.
These models can be realized by demanding DM interactions being
leptophilic,
namely, mediating DM particles only interact with the standard model
leptons.
The mediating particle does not have to be
light. The only constraint on the mass comes from the requirement of
sufficient enhancement of the DM annihilation cross section. In the model
with Sommerfeld enhancement the force-carrier particle should have a
mass $\sim m_{\rm dm}/E_{\rm F}$, while in the model with Brei-Wigner
enhancement the force-carrier particle should have a mass $\sim 2
m_{\rm dm}$\cite{bi-he}.

The DM annihilation models, although most intensively discussed in
the literature, are now facing strong observational and theoretical
constraints.
If the required boost factor ${\rm E_{\rm F}} \geq 10^{3}$ is not due
to the DM substructures, and if one assumes
such a boost and adopts a universal Navarro-Frenk-White (NFW)
density profile of DM halos, a large annihilation $\gamma-$ray flux
from nearby galaxy clusters is predicted.  With the parameters that
well fit the PAMELA and Fermi data\cite{Bergstrom09PRL} and the
standard assumptions for the limiting mass of the substructures within DM
halos, it is found that the EGRET upper limit in $\gamma$-ray flux is
violated in Virgo\cite{Pinzke09PRL}.
Similarly, the predicted gamma-ray flux violates the HESS upper limit
constraints from the Galactic center and some dwarf spheroidals, and
the predicted radio flux from synchrotron emission of the produced
$e^\pm$ from the Galactic Center also violates the current radio
observation upper limits.
The upper limits may be satisfied only if the DM
density profile is significantly less steep than the benchmark NFW and
Einasto profiles\cite{Bertone09JCAP}.  The diffuse $\gamma-$ray data
from the Fermi first year observations have imposed a very tight
constraint on the DM annihilation models and suggests
that the annihilation models likely only work in some ``fined-tuned''
situations\cite{Cirelli10,Papucci10}.
From theoretical point of view,
it is shown\cite{Feng09} that the enhancement required in the
Sommerfeld-enhanced DM annihilation models implies a thermal relic
density that is too small to allow the DM candidate suitable for
interpreting the $e^\pm$ feature to
account for all the DM in the universe. In Ref. \refcite{Feng09} the authors also derived an upper
bound on the possible Sommerfeld
enhancements from the observations of elliptical galactic DM halos and
showed that these bounds generically exclude such an explanation.

Counterarguments were also raised recently. It was argued\cite{Hutsi10}
that there are a lot of uncertainties in calculating the $\gamma$-ray
signals from the nearby dwarf spheroidals and the Galactic Center, and
that the annihilation models do not conflict with the Fermi gamma-ray
limits. It was also found that the annihilation models are better than
other models to account for the microwave signals collected by WMAP
in the Galactic Center\cite{Lin10}.

\subsubsection{Dark matter decay models}
The dark matter particles are often assumed to be perfectly stable as
the result of a symmetry, e.g. R-parity in the supersymmetric
models. However, from the gravitational evidence for the existence of
dark matter we can only infer that dark matter has to be
stable on timescales (much) longer than the age of the Universe. In
other words, DM particles are allowed to decay.
If they decay at a sufficiently high rate, the decay
products might be detectable as cosmic rays, which would result in
the observed $e^\pm$ excess.  There are in fact some
physically well-motivated dark matter candidates that have very
long lifetimes before decaying. For instance, gravitino dark matter,
which is unstable due to a small breaking of the R-parity,
constitutes an interesting scenario that leads to a thermal history
of the universe consistent with the observed abundances of primordial
elements, the observed dark matter relic abundance, and the observed
baryon asymmetry\cite{Buchmuller07JHEP}. Some other decaying DM
candidates discussed in the literature include
hidden gauge bosons\cite{decay}, hidden gauginos\cite{decay3},
right-handed sneutrinos in R-parity breaking scenarios\cite{decay1},
KK dark matter in a simple extension of the minimal universal extra
dimension models by introducing a small curvature\cite{decay25}, and
baryonic bound states of messenger quarks\cite{decay2}.  The late
decay of dark matter particles, for example gravitinos, can
produce gamma rays\cite{Ibarra08PRL,decay20}, positrons
\cite{Ibarra08JCAP}, neutrinos\cite{Covi09JCAP} and/or
antiprotons\cite{Ibarra08JCAP}, which contribute to the total
fluxes received at the Earth.

In the dark matter decay scenario(s), the source term of the
electrons/positrons is given by
\begin{equation}
q(\vec{r},p)={1\over \tau_{\rm dec}}{\rho_{\rm dm,\odot}(\vec{r})\over
m_{\rm dm}} {d{\cal N}_{\rm e^\pm} \over d{\cal E}},
\label{eq:Q_decay}
\end{equation}
where ${d{\cal N}_{\rm e^\pm} \over d{\cal E}}$ is the
energy spectrum of the electrons/positrons produced by the decay of
each DM particle. Evidently, in the decay model the parameters
involved regarding the DM properties are the mass $m_{\rm dm}$,
the lifetime $\tau_{\rm dec}$, and the spectrum of decay,
${d{\cal N}_{\rm e^\pm} \over d{\cal E}}$.

In order to account for the excess in the electron/positron channel but
not in the antiproton channel, the decay of DM should mainly produce
leptons with suppressed hadronic branching ratios. As a result,
the leptophilic decaying DM scenarios have attracted considerable
attention. As in the annihilation scenarios, leptophilic decaying DM
models can be broadly divided into two categories\cite{Chen10JCAP}.
One is that DM first decays into some new
light particles, which subsequently decay into muons or electrons,
while decays into hadrons are forbidden by kinematics. The other
is that the DM particle couples mainly to leptons due to
symmetry\cite{decay1} or geometric setup\cite{decay25}.

Although the decay models are not as widely investigated as the
annihilating models, they have some advantages that are not shared
by the annihilation models. First, in the
decay models there is no need of the ``boost factor'' given the
freedom of choosing the decay lifetime of the DM particles.
Second,
decay models suffer much less constraints from the gamma-ray and
radio observations from the Galactic
Center\cite{Liu10,Chen10JCAP,Cirelli10}.
Similar to the annihilation models, only the TeV $e^\pm$ pairs
generated within a distance $\sim 1$ kpc can reach us.
In the popular NFW profile of the dark matter halo with
$\rho(r)\propto r^{-1}$,
one expects a much higher DM density near the Galactic Center.
The gamma-ray signals produced in inverse Compton scattering
off the star light and CMB by pairs produced in the Galactic Center
can reach the Earth. Since the decay flux is proportional to the DM
density $\rho$ rather than proportional to $\rho^2$ as in the
annihilation models, the decay models predict a much fainter signal
than the annihilation models. The non-detection of such signals
pose strong constraints on the annihilation models, but not the
decay models (c.f. Ref.\refcite{Hooper09PRD-c,Hutsi10}).

\begin{figure}[t]
  \begin{center}
    \includegraphics[width=0.98\linewidth]{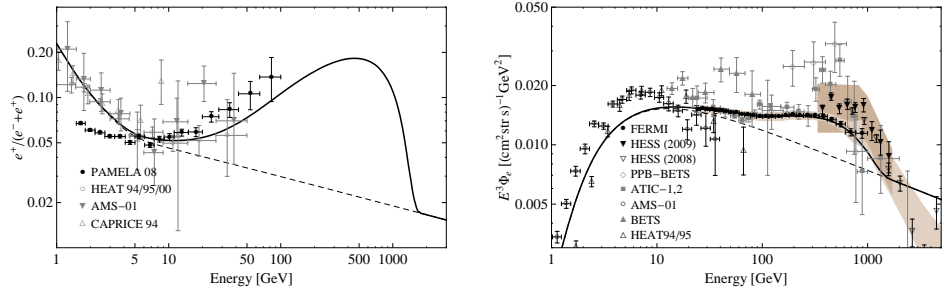}
  \end{center}
  \caption{Positron fraction (\textit{left panel}) and electron+positron flux
  (\textit{right panel}) for the fermionic DM decay ($\psi_{\rm dm}\rightarrow
  \mu^{+}\mu^{-}\nu$) model. The parameters are $m_{\rm dm} \sim 3.5$
  TeV and $\tau_{\rm dec} \sim 1.1\times 10^{26}$ sec. The dashed line
  shows the astrophysical background. A scalar dark matter particle
  decay ($\phi_{\rm dm}\rightarrow \mu^{+}\mu^{-}$) model
gives an equally well fit to the data for $m_{\rm dm} \sim 2.5$ TeV and
$\tau_{\rm dec} \sim 1.8\times 10^{26}$ sec.
The figures are from Ref.337.}
  \label{fig:decay1}
\end{figure}

\begin{figure}[t]
\begin{center}
\includegraphics[width=0.67\linewidth]{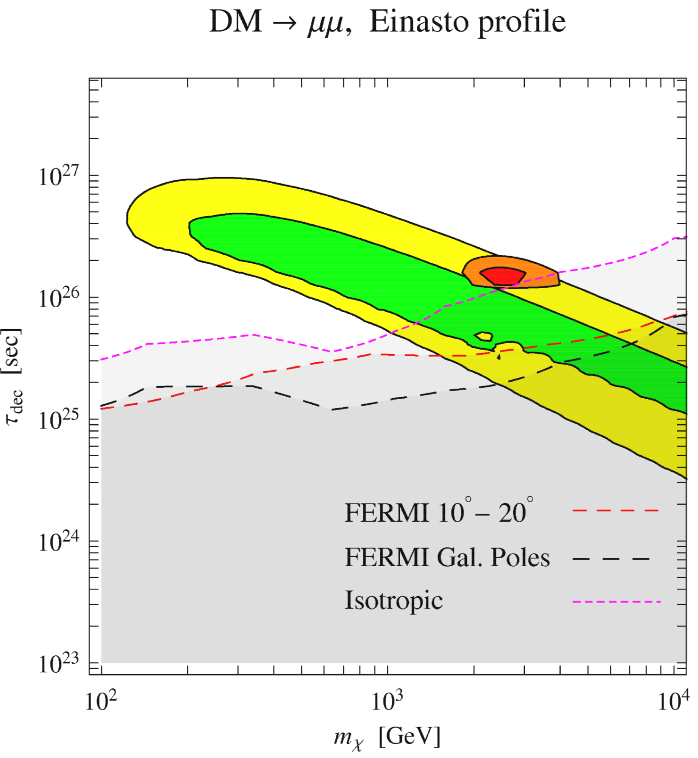}
\end{center}
\caption{The dark matter model parameter space
  $m_\chi - \tau_{\rm dec}$ (dark matter mass$-$lifetime)
  and the constraints from the diffuse
  galactic gamma ray measurements by the Fermi satellite in its first
  year operation (from Ref.326).
The exclusion contours
  are due to the Fermi observations of the `$10^\circ - 20^\circ$ `strip',
  the $|b| > 60^\circ$ `Galactic Poles' region, and the isotropic flux,
  respectively. The regions allowed to fit the PAMELA positron data
  (green and yellow bands, 95\%  and 99.999\% C.L. regions) and the
  PAMELA positron + Fermi and HESS data (red and orange blobs, 95\%
  and 99.999\% C.L. regions) in terms of decaying Dark Matter are also
  plotted. }
\label{fig:exclusiondecay}
\end{figure}

In the phenomenological approach, it is usually assumed that the dark
matter particle is either a fermion ($\psi_{\rm dm}$) or a scalar
($\phi_{\rm dm}$). Predictions for the positron
fraction for various decay channels and different dark matter masses
and lifetimes can be then computed\cite{decay3,Ibarra10JCAP}.
In the case of a
fermionic dark matter particle, people usually consider the two-body
decay channels $\psi_{\rm dm} \rightarrow Z^{0}\nu$, $\psi_{\rm dm}
\rightarrow W^{\pm}\ell^{\mp}$, as well as the three-body decay
channels $\psi_{\rm dm} \rightarrow \ell^{+}\ell^{-}\nu$, with
$\ell=e,~\mu,~\tau$ being the charged leptons. For
a scalar dark matter particle, one usually considers the two-body
decay channels $\phi_{\rm dm}\rightarrow Z^{0}Z^{0}$, $\phi_{\rm dm}
\rightarrow W^{+}W^{-}$, and $\phi_{\rm dm} \rightarrow
\ell^{+}\ell^{-}$. In light of the PAMELA and Fermi data, some
specific channels are found better than the others\cite{Ibarra10JCAP}.
One example of spectral modeling is shown in Fig.\ref{fig:decay1}.
The latest diffuse $\gamma-$ray data from Fermi first year
observations excludes the dark matter decay model unless the dark
matter decays mostly or even exclusively in
$\mu^{+}\mu^{-}$\cite{Cirelli10} (Fig.\ref{fig:exclusiondecay}, see
however Refs. \refcite{Papucci10,Zhang10} for looser constraints).

\section{Discussions and Conclusions}
After $\sim$ 40 years of efforts, the cosmic ray electron/positron
spectrum has been measured up to $\sim$ 1 TeV/$\sim 100$ GeV with an
unprecedented accuracy. The rising of the positron-to-electron ratio
above $\sim 10$ GeV, a tendency discovered in the
1970s\cite{Buffingston75}, has been confirmed by the PAMELA
satellite\cite{pamela-e} with high confidence.  The electron $+$
positron spectrum hardens above a few hundred GeV, as
discovered by ATIC\cite{atic}, PPB-BETS\cite{ppb-bets},
Fermi-LAT\cite{fermi} and HESS\cite{hess,hess1}. The rising
positron-to-electron ratio and the hardening electron $+$ positron
spectrum, unexpected in the standard cosmic ray model, are consistent
with each other and are called $e^{\pm}$ ``excesses''.  The detailed
electron $+$ positron spectral shape recorded by ATIC/PPB-BETS and
Fermi/HESS are however different. Future more advanced experimental
observations will pin down the shape of the spectrum more accurately.

A lot of efforts have been made to interpret the observed
eletron/positron excesses.  We summarize them in three main
categories:
\begin{itemize}
\item In the standard cosmic ray model, the
supernova remnants are the sole source of primary cosmic rays up to
PeVs.  The inhomogeneity of the supernova
remnants\cite{Shaviv09,Piran09} and/or the secondary particle
acceleration in supernova remnants\cite{Blasi09PRL,Ahlers09PRD}
may account for the PAMELA/ATIC/Fermi data. More speculatively,
some physical processes taking place in young
supernova remnants, for example, suppression of the high
energy $e^\pm$ cooling in the Klein-Nishina
regime\cite{Stawarz09ApJ,Schlickeiser09AA} and $e^\pm$ pair
production by interactions between high-energy CRs and background
photons\cite{Hu09ApJL}, may also account for the observed excesses.
\item It is possible that the observed $e^\pm$ excesses
originate from a new astrophysical CR source.
The leading candidate is pulsars\cite{Shen70,Aharonian95AA,Profumo09}.
Pulsars are known to be able to produce electron/positron pairs,
which may be accelerated to the desired high energies outside the
magnetospheres (likely in SNR shocks). No excess in
the spectrum of antiproton cosmic rays is expected, in agreement with
the PAMELA result. The escape of these pairs from the associated SNR,
however, is unclear. One popular assumption is that pulsars should
be mature, for which the associated SNR is too old to trap the
pairs. The potential problem is that mature pulsars have
lost a significant fraction of their initial kinetic energy and may
not be powerful enough to reproduce the observations. Some other
new astrophysical sources
discussed in the literature include microqusars\cite{Heinz02AA}
and local Gamma-ray Bursts\cite{Ioka10}.
The results are rather uncertain in these scenarios.
\item The most extensively discussed interpretation of the $e^\pm$
excesses is the annihilations or decays of dark matter. The excesses
in positrons and antiprotons in the CR spectra are two
``standard'' predictions of these models\cite{Jungman96PR,Bertone05PR}.
The detection of a rising positron-to-electron ratio is a good news
for these models. The non-detections of an excess in the antiproton
spectrum, however, calls for modifications of
most canonical dark matter models. Phenomenologically, one
usually assumes that\cite{arkani,leptophilic,He09IJPA,Chen10JCAP}
(1) the dark matter particles annihilate or decay into new particles,
which are so light that can only finally decay into $e^{\pm}$ and/or
$\mu^{\pm}$; (2) the DM is leptophilic and couples only to leptons
due to symmetry. In the DM annihilation models an unnatural large
``boost factor'' $\sim 10^{3}$ is needed to reproduce the
PAMELA/ATIC/Fermi detected $e^\pm$ fluxes. Such
an unusually large boost factor is difficult to interpret simply
by invoking substructures in the DM halo. Some
particle-physics-oriented suggestions, such as the Sommerfeld
and Briet-Wigner mechanisms, have been invoked to account for the
large excess. Many specific models have emerged and are
able to fit the electron/positron excesses. There are however some
additional observational constraints. For example, almost all dark
matter models predict a certain level excess of $\gamma$ ray. The
latest diffuse $\gamma-$ray data from the Fermi first year
observations have excluded a large portion of the DM parameter
space\cite{Cirelli10,Papucci10,Zhang10}. (In contrast, the
constraints on the astrophysical models are rather weak, since
they are ``local'' phenomena and just occupy a small fraction of
the volume of the Galaxy). In the case of annihilating DM, those
models generating predominantly $\mu^{\pm}$ and assuming a cored
Isothermal-like DM profile can fit the data with a mass
$m_{\rm dm}\sim 2$ TeV, but for other parameters, the model
suffers various observational constraints and is less favored
(c.f. Ref.\refcite{Hutsi10}). In the case of decaying DM, again
except those models generating predominantly $\mu^{\pm}$ (which can
fit the data with $m_{\rm dm}\sim 3$ TeV and $\tau_{\rm dec} \sim
2\times 10^{26}$ sec), most models are not favored by various
observational constraints.
It is worth noting that recently direct searches of
dark matter have been performed by various experimental groups.
For example, the CDMS II collaboration has
observed two candidate events\cite{CDMS-II}. If interpreted as DM
recoil on nuclei, the DM particles detected by CDMS II should be
relatively light, with a mass $m_{\rm dm} \sim {\cal O}(100)$ GeV.
This is much lower than that needed
in the $e^\pm$ excesses modeling. Due to low statistics,
it is of course too early to say that the observed $e^\pm$ excesses
are not of the DM origin.
In any case, direct detection experiments and the high
energy colliders (such as the Large Hadron Collider)
would play an essential role to
probe the nature of DM, and to provide a test to
the current DM interpretations of the $e^\pm$
excesses. In the near future, neutrino observations by the
upcoming high energy neutrino detectors (such as IceCube)
would also help to constrain the DM nature of the
$e^\pm$ excesses\cite{Bertone05PR,Hooper:2007qk}.
\end{itemize}

All the models reviewed here assume that the detected
cosmic ray $e^\pm$ excesses are due to one physical
process. In reality, this may not be the case.
For example, it is possibly that the detected excesses
are mainly contributed by astrophysical processes, and the DM
annihilation/decay signals are outshone. If so, the extraction of the
DM properties via indirect detection experiments would be even more
challenging.  Nevertheless further indirect
detection experiments are still necessary since they are complementary
to the direct detection experiments and the collider experiments for
revealing the properties of DM. For instance, collider experiments may
identify a long-lived, weakly interacting particle and measure
its mass. On the other hand, it would be difficult
to test its cosmological abundance and stability. Such goals, however,
are achievable in the direct and/or indirect detection experiments.

We conclude that both astrophysical processes and DM
annihilation or decay models can provide a self-consistent explanation
to the recently observed $e^\pm$ excesses in the cosmic ray spectrum.
At present the dark matter interpretations
seem to be more exotic than some astrophysical scenarios.
This is in particular the case for the dark matter
annihilation models which are tightly constrained by the
$\gamma-$ray observational data by Fermi.  In fact the latest
diffuse gamma-ray data have excluded a large region of the DM
parameter space that is favored in interpreting the $e^\pm$ excesses.
More experimental observations are needed to pin down the nature of
DM and its implications for the $e^\pm$ excess.
Among the promising astrophysical models, the inhomogeneous
distribution of supernova remnants has a smoking gun prediction
that the positron-to-electron ratio is expected to decrease above an
energy $\sim 100$ GeV. The pulsar model, instead, predicts an increasing
positron-to-electron ratio up to the energy $\sim 1$ TeV. These two
predictions will soon be tested by the ongoing PAMELA and the upcoming
AMS-2 experiments. Both the inhomogeneous SNR model and the pulsar model
predict a possible wiggle-like structure in the TeV energy range,
which may be constrained by the observations of H.E.S.S-like ground
based Cherenkov telescopes.

\medskip
Acknowledgements:
This work is supported in part by the National basic research program of China under grant
2009CB824800 and a special grant of Chinese
Academy of Sciences (for Y.Z.F), and by NASA NNX09AT66G, NNX10AD48G,
and NSF AST-0908362 (for B.Z.)
We thank  P. H. Tam, Q. Yuan, X. J. Bi, and H. B. Hu for communications, and also D. Hooper for reading the manuscript.


\begin{thebibliography}{0}

\bibitem{Ginzburg96PhysU} V.~L.~Ginzburg, {\it Physics-Uspekhi.} {\bf 39}, 155 (1996).

\bibitem{Elster1901} J.~Elster and H. Geitel, {\it Phys. Zs.} {\bf 2}, 560 (1901).

\bibitem{Wilson1901} C. T. R. Wilson, {\it Camb. Phys. Soc. Proc.} {\bf 11}, 32 (1901).

\bibitem{Wilson1901b} C. T. R. Wilson, {\it Roy. Soc. Proc.} {\bf 68}, 151 (1901).

\bibitem{Hess1912} V. F. Hess, {\it Phys. Zs.} {\bf 13}, 1084 (1912).

\bibitem{Kolhorster1913} W. Kolhorster, {\it Phys. Zs.} {\bf 14}, 1066 (1913).

\bibitem{Ginzburg64book} V. L. Ginzburg and S. I. Syrovatskii, The Origin of Cosmic Rays. New York: Macmillan
(1964)

\bibitem{Meyer69ARAA} P. Meyer, \araa {\bf 7}, 1 (1969).

\bibitem{Hillas84ARAA} P. Hillas, \araa {\bf 22}, 425 (1984).

\bibitem{Nagano00RMP} M. Nagano and A. A. Watson, {\it Reviews of Modern Physics.} {\bf 72}, 689 (2000).

\bibitem{Earl61PRL} J. A. Earl, \PRL {\bf 6}, 125 (1961).

\bibitem{Meyer61PRL} P. Meyer and R. Vogt, \PRL {\bf 6}, 193 (1961).

\bibitem{De-Shong64PRL} J. A. De Shong, R. H. Hildebrand and P. Meyer, \PRL {\bf 12}, 3 (1964).

\bibitem{Strong07Rev} A. W. Strong, I. V. Moskalenko and
 V. S. Ptuskin, {\it Annu. Rev. Nucl. Part. Sci.} {\bf 57}, 285 (2007).

\bibitem{Buffingston74PRL} A. Buffingston, C. D. Orth and G. F. Smoot, \PRL {\bf 33}, 34 (1974).

\bibitem{Daniel65PRL} R. R. Daniel and S. A. Stephens, \PRL {\bf 15}, 769 (1965).

\bibitem{Daniel70SSRev} R. R. Daniel and S. A. Stephens, {\it Space Science Reviews.} {\bf 10}, 599 (1970).

\bibitem{Agrinier69} B. Agrinier, {\it et al.}, {\it Lett. Nuovo Cimento.} {\bf 1}, 53 (1969).

\bibitem{Muller87} D. M\"{u}ller and K. K. Tang, \apj {\bf 312}, 183 (1987).

\bibitem{Golden87AA} R. L. Golden, {\it et al.}, \aaa  {\bf 188}, 145 (1987).

\bibitem{pamela-e}
  O.~Adriani {\it et al.}  [PAMELA Collaboration],
  \Nature {\bf 458}, 607 (2009).

\bibitem{pamela-e1}
  O.~Adriani {\it et al.}  [PAMELA Collaboration],
  \apj submitted, arXiv:1001.3522 (2010).

\bibitem{Protheroe82} R. J. Protheroe, \apj {\bf 254}, 391 (1982).

\bibitem{atic}
  J.~Chang {\it et al.} [ATIC Collaboration],
  \Nature {\bf 456}, 362 (2008).

\bibitem{ppb-bets}
  S.~Torii {\it et al.},
  arXiv:0809.0760 [astro-ph].

\bibitem{fermi}
  A.~A.~Abdo {\it et al.}  [The Fermi LAT Collaboration],
 \PRL  {\bf 102}, 181101 (2009).

\bibitem{He09IJPA} X. G. He, {\it Modern Physics Letters A.} {\bf 24}, 2139 (2009).

\bibitem{Boezio09NJP} M. Boezio, {\it et al.}, {\it New Journal of Physics.} {\bf 11}, 105023 (2009).

\bibitem{Critchfield52PRev} C. L. Critchfield, E. P. Ney and S. Oleksa, {\it Phys. Rev.} {\bf 85}, 461 (1952)

\bibitem{Agrinier64PRL} B. Agrinier, Y. Koechlin and B. Parlier, \PRL {\bf 13}, 377 (1964).

\bibitem{Hartman67ApJ} R. C. Hartman, \apj {\bf 150}, 371 (1967).

\bibitem{Fanselow69ApJ} J. L. Fanselow, {\it et al.}, \apj {\bf 158}, 771 (1969).

\bibitem{Golden94ApJ} R. L. Golden, {\it et al.}, \apj {\bf 436}, 769 (1994).

\bibitem{Barbiellini96AA} G. Barbiellini, {\it et al.}, \aaa {\bf 309}, L15 (1996).

\bibitem{Golden96ApJ} R. L. Golden, {\it et al.}, \apj {\bf 457}, L103 (1996).

\bibitem{Barwick97ApJ} S. W. Barwick, {\it et al.}, \apj {\bf 482}, L191 (1997);
 J. J. Beatty, {\it et al.}, \PRL, {\bf 93}, 241102 (2004).

\bibitem{Boezio01ApJ} M. Boezio, {\it et al.}, \apj {\bf 532}, 653 (2001).

\bibitem{Torii01ApJ} J. J. Torii, {\it et al.}, \apj {\bf 559}, 973 (2001).

\bibitem{Cline64PRL} T. L. Cline, G. H. Ludwig and F. B. McDonald, \PRL {\bf 13}, 786 (1964).

\bibitem{Fan68ApJ} C. Y. Fan {\it et al.}, \apj {\bf 151}, 737 (1968).

\bibitem{Simnent69ApJ} G. M. Simnent and F. B. McDonald, \apj {\bf 157}, 1435 (1969).

\bibitem{Hurford74ApJ} G. J. Hurford, {\it et al.}, \apj {\bf 192}, 541 (1974).

\bibitem{ams}
  M.~Aguilar {\it et al.}  [AMS-01 Collaboration],
 \PLB {\bf 646}, 145 (2007).

\bibitem{hess}
 F. Aharonian et al. [H.E.S.S. Collaboration], \PRL {\bf 101}, 261104 (2008).

\bibitem{Anand68PRL}  K. C. Anand, R. R. Daniel and S. A. Stephens, \PRL {\bf 20} 764 (1968)

\bibitem{Muller73ApJ} D. M\"{u}ller and P. Meyer, \apj {\bf 186}, 841 (1973).

\bibitem{Hartmann77PRL} G. Hartmann, D. M\"{u}ller and T. Prince, \PRL {\bf 38}, 1368 (1977).

\bibitem{Tang84ApJ} K. K. Tang, \apj {\bf 278}, 881 (1984).

\bibitem{Buffingston75} A. Buffingston, C. D. Orth and G. F. Smoot, \apj {\bf 199}, 669 (1975).

\bibitem{Nishimura80ApJ} J. Nishimura, {\it et al.}, \apj {\bf 238}, 394 (1980).

\bibitem{Kobayashi99ICRC} T. Kobayashi, {\it et al.}, {\it Proceedings of the 26th International Cosmic Ray Conference.} {Vol. 3}, 61 (1999).

\bibitem{Anand73ICRC} K. C. Anand, R. R. Daniel and S. A. Stephens, {\it Proceedings of the 13th International Cosmic Ray Conference.} {Vol. 1}, 355 (1973).

\bibitem{Earl72JGR} J. A. Earl, D. E. Neely and T. A. Rygg, {\it J. Geophys. Res.} {\bf 77}, 1087 (1972).

\bibitem{Meegan75ApJ} C. A. Meegan and J. A. Earl, \apj {\bf 197}, 219 (1975).

\bibitem{Beatty04PRL} J. J. Beatty, {\it et al.}, \PRL {\bf 93}, 241102 (2004).

\bibitem{Boezio01ASR} M. Boezio, {\it et al.}, {\it Adv. Spac. Res.} {\bf 27}, 669 (2001).

\bibitem{pamela-p}
  O.~Adriani {\it et al.},
    \PRL  {\bf 102}, 051101 (2009).

\bibitem{hess1}
F. Aharonian et al. [H.E.S.S. Collaboration], \aaa {\bf 508}, 561 (2009).

\bibitem{Strong98ApJ} A. W. Strong and I. V. Moskalenko, \apj {\bf 509}, 212
(1998).

\bibitem{Seo94AA} E. S. Seo and V. S. Ptuskin, \apj {\bf 431}, 705 (1994).

 \bibitem{Yin08PRD}
  P.~F.~Yin et al.,
 \PRD {\bf 79}, 023512 (2009).

  \bibitem{Kamionkowski91PRD}
  M. Kamionkowski and M. S. Turner, \PRD {\bf 43},
1774 (1991).

 \bibitem{Baltz98PRD}
E. A. Baltz and J. Edsjo, \PRD {\bf 59}, 023511 (1999).

\bibitem{Moskalenko98ApJ} I. V. Moskalenko and A. W. Strong, \apj
 {\bf 493}, 694 (1998).

\bibitem{Serpico09} P. D. Serpico, \PRD {\bf 79}, 021302 (2009).

\bibitem{Shaviv09} N. J. Shaviv, E. Nakar and T. Piran, \PRL {\bf 103}, 111302 (2009).

\bibitem{Piran09} T. Piran, N. J. Shaviv and E. Nakar, 2009 (arXiv:0905.0904).

\bibitem{Grasso09AP} D. Grasso, et al., {\it Astropart. Phys.} {\bf 32}, 140 (2009).

\bibitem{Jauch76book} J. M. Jauch and F. Rohrlich, The Theory of photons and Electrons, 2nd Edition, Springer-Verlag, New York (1976).

\bibitem{Erber66RMP} T. Erber, {\it Rev. Mod. Phys.} {\bf 38}, 626 (1966).

\bibitem{Gebauer09} I. Gebauer and W. de Boer, \aaa submitted (arXiv:0910.2027).

\bibitem{Aharonian95AA} F. A. Aharonian, A. M. Atoyan, and H. J. Voelk, \aaa {\bf 294}, L41 (1995).

\bibitem{Strong01ICRC} A. W. Strong and I. V. Moskalenko, {\it ICRC} {\bf 5}, 1964 (2001) [arXiv:astro-ph/0106505].

\bibitem{Kobayashi04ApJ} T. Kobayashi, Y. Komori, K. Yoshida, and J. Nishimura, \apj {\bf 601}, 340 (2004).


\bibitem{Hillas05} A. M. Hillas, {\it J. Phys. G: Nucl. Part. Phys.} {\bf 31}, R95 (2005).

\bibitem{Aharonian06AA} F. Aharonian, et al., \aaa {\bf 457}, 899 (2006).

\bibitem{Lacey01ApJ} C. K. Lacey and N. Duric, \apj {\bf 560}, 719
(2001).

\bibitem{Biermann09PRL} P. L. Biermann, J. K. Becker, A. Meli, W. Rhode, E. S. Seo and T. Stanev,
\PRL {\bf 103}, 061101 (2009).


\bibitem{Blasi09PRL} P. Blasi, \PRL {\bf 103}, 051104 (2009).

\bibitem{Ahlers09PRD} M. Ahlers, P. Mertsch and S. Sarkar, \PRD {\bf 80}, 123017 (2009).

\bibitem{Eichler80ApJ}
  D. Eichler, \apj {\bf 237}, 809 (1980).

\bibitem{Cowsik80ApJ}
R. Cowsik, \apj {\bf 241}, 1195 (1980).

\bibitem{Blasi09PRL-b} P. Blasi and P. D. Serpico, \PRL {\bf 103}, 081103
(2009).

\bibitem{Mertsch09PRL} P. Mertsch and S. Sarkar, \PRL {\bf 103}, 081104
(2009).



\bibitem{Blum70PRD} G. R. Blumenthal, \PRD {\bf 1}, 1596 (1970).

\bibitem{Hu09ApJL} H. B. Hu, Q. Yuan, B. Wang, C. Fan, J. L. Zhang, and X. J. Bi, \apj {\bf 700}, L170 (2009).

\bibitem{Blumenthal70RMP} G. B. Blumenthal and R. J. Gould, {\it Rev. Mod. Phys.} {\bf 42}, 237 (1970).

\bibitem{Schlickeiser09AA} R. Schlickeiser and J. Ruppel, {\it New. J. Phys.}, {\bf 12}, 033044 (2010).

\bibitem{Stawarz09ApJ} L. Stawarz, V. Petrosian and R. D. Blandford, \apj, {\bf 710}, 236 (2010).

\bibitem{Shen70} C. S. Shen, \apj {\bf 162}, L181 (1970).

\bibitem{Yuksel09PRL} H. Y\"{u}ksel, M. D. Kistler, and T. Stanev, \PRL {\bf 103}, 051101 (2009).

\bibitem{Hooper09JCAP} D. Hooper, P. Blasia, and P. D. Serpicoe, \JCAP {\bf 01}, 025 (2009).


\bibitem{Manchester77} R. N. Manchester, and J. H. Taylor, Pulsars, San Francisco: W. H. Freeman (1977)

\bibitem{Abdo09Sci} A. A. Abdo et al. [The Fermi Collaboration], {\it Science} {\bf 325}, 840  (2009).

\bibitem{Abdo09ApJ} A. A. Abdo et al. [The Fermi Collaboration], {\it Astrophys. J. Supp.} {\bf 187}, 460 (2010).


\bibitem{Ruderman75ApJ} M. A. Ruderman and P. G. Sutherland, \apj {\bf 196}, 51 (1975).

\bibitem{Usov96ApJ} V. V. Usov and D. B. Melrose, \apj {\bf 464}, 306 (1996).

\bibitem{Arons83ApJ} J. Arons, \apj, {\bf 266}, 215 (1983).

\bibitem{Daugherty96ApJ} J. K. Daugherty and A. K. Harding, \apj {\bf 458}, 278 (1996).

\bibitem{Zhang00ApJ} B. Zhang and A. K. Harding, \apj, {\bf 532}, 1150 (2000).

\bibitem{Cheng86ApJ} K. S. Cheng, C. Ho, and M. Ruderman, \apj {\bf 300}, 500 (1986).

\bibitem{Romani96ApJ} R. W. Romani, \apj {\bf 470}, 469 (1996).

\bibitem{Zhang97ApJ} L. Zhang and K. S. Cheng, {\bf 487}, 370 (1997).

\bibitem{Arons79ApJ} J. Arons and E. T. Scharlemann, \apj, {\bf 231}, 854 (1979).

\bibitem{Muslimov03ApJ} A. G. Muslimov, and A. K. Harding, \apj, {\bf 588}, 430 (2003).

\bibitem{Pacini67Nature} F. Pacini, {\it Nature} {\bf 216}, 567 (1967).

\bibitem{Gold68Nature} T. Gold, {\it Nature} {\bf 218}, 731 (1968).

\bibitem{Ostriker69ApJ} J. P. Ostriker and J. E. Gunn, \apj {\it 157}, 1395 (1969).

\bibitem{Malyshev09PRD} D. Malyshev, I. Cholis and J. Gelfand, \PRD {\bf 80}, 063005 (2009).

\bibitem{Profumo09} S. Profumo, arXiv:0812.4457 (2009).

\bibitem{Harding87ApJ} A. K. Harding and  R. Ramaty, {\it ICRC} {\bf 2}, 92 (1987).

\bibitem{Chi96ApJ} X. Chi, K. S. Cheng, and E. C. M. Young, \apj {\bf 459}, L83 (1996).

\bibitem{Zhang01AA} L. Zhang and K. S. Cheng, \aaa {\bf 368}, 1063 (2001).

\bibitem{Busching08} I. Busching, C. Venter, and O. C. de Jager, {\it Adv. Space Res.} {\bf 42}, 497 (2008).
\bibitem{Atoyan95PRD} A. M. Atoyan, F. A. Aharonian, and H. J. V\"olk, \PRD {\bf 52}, 3265 (1995).

\bibitem{Pohl09PRD} Pohl, M. 2009, \PRD {\bf 79}, 041301 (2009).

\bibitem{Mirabel99ARAA} I. F. Mirabel, L. F. Rodriguez, \araa {\bf 37}, 409 (1999).

\bibitem{Bosch-Ramon08IJMPD} V. Bosch-Ramon and
D. Khangulyan, {\it Int. J. Mod. Phys. D} {\bf 18}, 347 (2009).

\bibitem{Heinz02AA} S. Heinz and R. Sunyaev, \aaa {\bf 390}, 751 (2002).

\bibitem{Fender05MN} R. F. Fender, T. J. Maccarone, and Z. van. Kestern, \MNRAS {\bf 360}, 1085 (2005).

\bibitem{Albert06Science} J. Albert, et al., {\it Science} {\bf 312}, 1771 (2006).

\bibitem{Tavani09Nature} M. Tavani, et al., {\it Nature} {\bf 462}, 620 (2009).

\bibitem{Guessoum06AA} N. Guessoum, P. Jean, and N. Prantzos, \aaa, {\bf 457}, 753 (2006).

\bibitem{Klebesadel73ApJ} R. W. Klebesadel, I. B. Strong, and R. A. Olson, \apj,
{\bf 182}, L85 (1973).


\bibitem{Piran04RMP} T. Piran, {\it Rev. Mod. Phys.},
{\bf 76}, 1143 (2004).

\bibitem{Guetta05ApJ} D. Guetta, T. Piran and E. Waxman, \apj, {\bf 619}, 412 (2005).

\bibitem{Zhang04IJMPA} B. Zhang and P. M\'esz\'aros,  {\it Int. J. Mod. Phys. A}, {\bf 19}, 2385 (2004).

\bibitem{Frail01ApJ} D. Frail, et al. \apj, {\bf 562}, L55 (2001).

\bibitem{Coward05MN} D. M. Coward, \MNRAS, {\bf 360}, L77 (2005).

\bibitem{Liang07ApJ} E. Liang, B. Zhang, F. Virgili, Z. G. Dai, \apj, {\bf 662}, 1111 (2007).

\bibitem{Ioka10} K. Ioka, {\it Prog. Theo. Phys.} {\bf 123}, 743 (2010).

\bibitem{Fruchter06Nature} A. S. Fruchter, et al., {\it Nature.} {\bf 441}, 463 (2006).

\bibitem{Savaglio09ApJ} S. Savaglio, K. Glazebrook, D. Le Borgne, \apj {\bf 691}, 182 (2009).

\bibitem{Stanek06AcA} K. Z. Stanek et al. {\it Acta Astron.} {\bf 56}, 333 (2006)

\bibitem{DAvanzo10} P. D'Avanzo et al. \aaa, submitted (2010)

\bibitem{fp08} Y. Z. Fan and T. Piran, {\it Front. Phys. China.} {\bf 3}, 306 (2008).

\bibitem{Racusin08Nature} J. L. Racusin, et al., {\it Nature.} {\bf 455}, 183 (2008).

\bibitem{Jungman96PR} G. Jungman, M. Kamionkowski and K. Griest, {\it Phys. Rept.} {\bf 267}, 195
(1996).

\bibitem{Bergstrom00RPP} L. Bergstr\"{o}m, {\it Rep. Prog. Phys.} {\bf 63}, 793 (2000).

\bibitem{Munoz04IJMPA} C. Munoz, {\it Int. J. Mod. Phys. A.} {\bf 19}, 3093 (2004).

\bibitem{Bertone05PR} G. Bertone, D. Hooper and J. Silk, {\it Phys. Rept.} {\bf 405}, 279 (2005).

\bibitem{Hooper:2007qk} D.~Hooper and S.~Profumo, {\it Phys. Rept.}  {\bf 453}, 29 (2007).

\bibitem{Olive08ASR} K. A. Olive, {\it Advances in Space Research.} {\bf 42}, 581 (2008).

\bibitem{Steffen09EPJC} F. D. Steffen, {\it The European Physical Journal C.} {\bf 59}, 557 (2009).

\bibitem{Hooper09Rev} D. Hooper, (2009) [arXiv:0901.4090].

\bibitem{arXiv:0907.1912} G. D'Amico, M. Kamionkowski and K. Sigurdson, arXiv:0907.1912 (2009).

\bibitem{Bergstrom09NPJ} L. Bergstr\"{o}m, {\it New. J. Phys.} {\bf 11}, 105006 (2009).

\bibitem{Ellwanger09} U. Ellwanger, C. Hugonie, and A. M. Teixeira, arXiv:0910.1785 (2010).

\bibitem{Feng09ARAA} J. L. Feng, \araa, in press (arXiv:1003.0904)

\bibitem{Zwicky:1933} F. Zwicky, {\it Helv. Phys. Acta} {\bf 6}, 110 (1933).

\bibitem{Peacock} J. A. Peacock, {\it Cosmological Physics.}, Cambridge University Press, (1999).

\bibitem{Iocco:va08} F. Iocco, G. Mangano, G. Miele, O. Pisanti, and P. D. Serpico, {\it Phys. Rept.} {\bf 472}, 1 (2009).

\bibitem{Begeman91MN} K. G. Begeman, A. H. Broeils, and R. H. Sanders, \MNRAS {\bf 249}, 523 (1991).

\bibitem{Blandford92ARAA} R. D. Blandford and R. Narayan, \araa {\bf 30}, 311 (1992).

\bibitem{Bahcall98ApJ} N. Bahcall and X. Fan, \apj {\bf 504}, 1 (1998).

\bibitem{Spergel03} D. N. Spergel et al. {\em Astrophys. J. Supp.}  {\bf 148}, 175 (2003).

\bibitem{Riess98} A. G. Riess et al. {\em Astron. J.} {\bf 116}, 1009 (1998).

\bibitem{Komatsu10} E. Komatsu et al. {\em Astrophys. J.} submitted (2010) (arXiv:1001.4538).

\bibitem{Eisenstein05} D. J. Eisenstein et al. \apj {\bf 633}, 560 (2005).

\bibitem{Alcock00} C. Alcock et al. \apj {\bf 542}, 281 (2000).






\bibitem{unparticles}
  T.~Kikuchi and N.~Okada,
\PLB {\bf 665}, 186 (2008).

\bibitem{qballs}
  A.~Kusenko and M.~E.~Shaposhnikov,
  \PLB {\bf 418}, 46 (1998).

\bibitem{LEP03PLB} ALEPH, DELPHI, L3, and OPAL Collaborations, {\it
    Phys. Lett.} {\bf B565}, 61 (2003)

\bibitem{LEP} ALEPH, DELPHI, L3, and OPAL Collaborations,
http://lepewwg.web.cern.ch/LEPEWWG

\bibitem{wess} J. Wess and B. Zumino, \NPB {\bf 70}, 39 (1974).

\bibitem{fayet} P. Fayet, \PLB {\bf 86}, 272 (1979).

\bibitem{ellis0} J. Ellis, J.S. Hagelin, D.V. Nanopoulos, K. Olive,
M. Srednicki, \NPB {\bf 238}, 453 (1984).

\bibitem{haberkane} H.E. Haber and G.L. Kane, {\it Phys. Rep.} {\bf 117}, 75 (1985).

\bibitem{Falk94PLB} T. Falk, K.A. Olive, and M. Srednicki, \PLB {\bf 339}, 248 (1994).

\bibitem{Arina07JHEP} C. Arina and N. Fornengo, {\it JHEP} {\bf 11}, 029 (2007).

\bibitem{Feng03PRL} J. L. Feng, A. Rajaraman, and F. Takayama, \PRL {\bf 91}, 011302 (2003).

\bibitem{Moroi93PLB} T. Moroi, H. Murayama, and M. Yamaguchi, \PLB {\bf 303}, 289 (1993).

\bibitem{Buchmuller07JHEP} W. Buchmuller, L. Covi, K. Hamaguchi, A. Ibarra, and T. Yanagida, {\it JHEP} {\bf 0703}, 037 (2007).

\bibitem{Buchmuller09JCAP} W. Buchmuller, A. Ibarra, T. Shindou, F. Takayama, and D. Tran, \JCAP {\bf 0909}, 021 (2009).


\bibitem{PQ} R. Peccei and H.R. Quinn, \PRL
{\bf 38}, 1440 (1977).

\bibitem{Kim09RMP} J. E. Kim and G. Carosi, \RMP {\bf 82}, 557 (2010).

\bibitem{ma}
  N.~G.~Deshpande and E.~Ma,
  \PRD {\bf 18}, 2574 (1978).

\bibitem{ma-2}
  E.~Ma,
  \PRD {\bf 73}, 077301 (2006).

\bibitem{rychkov}
  R.~Barbieri, L.~J.~Hall and V.~S.~Rychkov,
 \PRD {\bf 74}, 015007 (2006).

\bibitem{honorezlopez}
  L.~Lopez Honorez, E.~Nezri, J.~L.~Oliver and M.~H.~G.~Tytgat,
  \JCAP {\bf 0702}, 028 (2007).

\bibitem{Dolle09PRD} E. Dolle and S. F. Su, \PRD {\bf 80}, 055012 (2009).

\bibitem{Gustafsson07PRL} M. Gustafsson, E. Lundstr\"{o}m, L. Bergstr\"{o}m, and J. Edsj\"{0}, \PRL
{\bf 99}, 041301 (2007).

\bibitem{Goto92PLB} T. Goto and M. Yamaguchi, \PLB {\bf 276}, 103 (1992).

\bibitem{Bonometto94PRD} S. A. Bonometto, F. Gabbiani, and A. Masiero, \PRD {\bf 49}, 3918 (1994).

\bibitem{Covi99PRL} L. Covi, J.E. Kim, and L. Roszkowski, \PRL {\bf 82}, 4180 (1999).

\bibitem{Chun00PRD} E. J. Chun, H. B. Kim, and D. H. Lyth, \PRD {\bf 62}, 125001 (2000).

\bibitem{Covi09NJP} L. Covi and J.E. Kim, {\it New. J. Phys.} {\bf 11}, 105003 (2009).

\bibitem{hut} P. Hut, {\it Phys. Lett.} {\bf 69B}, 85 (1977).

\bibitem{leeweinberg} B.W. Lee and S. Weinberg, \PRL {\bf 39}, 165 (1977).

\bibitem{vysotsky} M.I. Vysotsky, A.D. Dolgov and Ya. B. Zel'dovich,
{\it JETP Lett.} {\bf 26}, 188 (1977).

\bibitem{gunn} J. E. Gunn, B.W. Lee, I. Lerche, D.N. Schramm and
G. Steigman, \apj {\bf 223}, 1015 (1978).

\bibitem{Kraus05} Ch. Kraus et al. {\it Euro. Phys. J. C.} {\bf 40}, 447 (2005)

\bibitem{tg} S. Tremaine and J.G. Gunn, \PRL {\bf 42}, 407 (1979).

\bibitem{sterile}
  S.~Dodelson and L.~M.~Widrow,
\PRL  {\bf 72}, 17 (1994).

\bibitem{Shi98PRL}
  X.~D.~Shi and G.~M.~Fuller,
  \PRL  {\bf 82}, 2832 (1999).


\bibitem{abazajian}
  K.~N.~Abazajian,
  arXiv:0903.2040 (2010).

\bibitem{shaposh}
  A.~Boyarsky, O.~Ruchayskiy and M.~Shaposhnikov,
 {\it Ann.Rev.Nucl.Part.Sci.} {\bf 59}, 191 (2009).

\bibitem{Abazajian01PRD} K. Abazajian, G. M. Fuller, and M. Patel, \PRD {\bf 64}, 023501 (2001).

\bibitem{Boyarsky09PRL}
A. Boyarsky, J. Lesgourgues, O. Ruchayskiy, and M. Viel, \PRL {\bf 102}, 201304 (2009).


\bibitem{Cheng02PRD1}
H.~C.~Cheng, K.~T.~Matchev and M.~Schmaltz,
\PRD {\bf 66}, 036005 (2002).

\bibitem{Servant03NPB}
G.~Servant and T.~M.~P.~Tait,
\NPB {\bf 650}, 391 (2003).

\bibitem{Cirelli09NPB} M. Cirelli and A. Strumia, \NPB {\bf 821}, 399 (2009).

\bibitem{Barger09PLB} V. Barger, W.Y. Keung, D. Marfatia, and G. Shaughnessy,
\PLB {\bf 672}, 141 (2009).

\bibitem{Cholis09PRD} I. Cholis, L. Goodenough, D. Hooper, M. Simet, and N.
Weiner, \PRD {\bf 79}, 123505 (2009).


\bibitem{Cirelli09NPB-1} M. Cirelli, M. Kadastik, M. Raidal, and A. Strumia,
\NPB {\bf 813}, 1 (2009).

\bibitem{Bergstrom08PRD} L. Bergstrom, T. Bringmann, and J. Edsjo, Phys. Rev. D
78, 103520 (2008).

\bibitem{arkani} N. Arkani-Hamed, D. P. Finkbeiner, T. Slatyer, and N.
Weiner, \PRD {\bf 79}, 015014 (2009).

\bibitem{Pospelov09PLB} M. Pospelov and A. Ritz, \PLB {\bf 671}, 391 (2009).

\bibitem{Cholis09JCAP} I. Cholis, D. P. Finkbeiner, L. Goodenough, and N.Weiner, \JCAP
{\bf 0912}, 007 (2009).

\bibitem{Dicus77PRL} D. A. Dicus, E. N. Kolb, and V. L. Teplitz, \PRL {\bf 39}, 168 (1977).

\bibitem{Tylka89PRL} A.J. Tylka, \PRL {\bf 63}, 840 (1989).

\bibitem{Turner90PRD} M. S. Turner and F. Wilczek, \PRD {\bf 42}, 1001 (1990).

\bibitem{Baltz99PRD} E.A. Baltz and J. Edsjo, \PRD {\bf 59}, 023511 (1999).

\bibitem{leptophilic}
  P.~J.~Fox and E.~Poppitz,
  Phys.\ Rev.\  D {\bf 79}, 083528 (2009).

\bibitem{lep1}
  S.~Baek and P.~Ko,
  \JCAP {\bf 0910}, 011 (2009).

\bibitem{lep2}
  H.~S.~Goh, L.~J.~Hall and P.~Kumar,
  {\it JHEP} {\bf 0905}, 097 (2009).

\bibitem{lep3}
  H.~Davoudiasl,
 \PRD {\bf 80}, 043502 (2009).

\bibitem{bi-he}
  X.~J.~Bi, X.~G.~He, and Q.~Yuan,
  \PLB {\bf 678}, 168 (2009).

\bibitem{susy1}
  J.~Hisano, M.~Kawasaki, K.~Kohri and K.~Nakayama,
  Phys.\ Rev.\  D {\bf 79}, 063514 (2009).

\bibitem{susy2}
  K.~Ishiwata, S.~Matsumoto and T.~Moroi,
  Phys.\ Lett.\  B {\bf 675}, 446 (2009).

\bibitem{susy3}
  J.~Kalinowski, S.~F.~King and J.~P.~Roberts,
  JHEP {\bf 0901}, 066 (2009).

\bibitem{susy4}
  R.~Allahverdi, B.~Dutta, K.~Richardson-McDaniel and Y.~Santoso,
  Phys.\ Rev.\  D {\bf 79}, 075005 (2009).

\bibitem{susy5}
  P.~Grajek et al.,
\PRD {\bf 79}, 043506 (2009).

\bibitem{susy6}
  J.~H.~Huh, J.~E.~Kim and B.~Kyae,
  arXiv:0812.5004 [hep-ph].

\bibitem{susy7}
  I.~Gogoladze, R.~Khalid, Q.~Shafi and H.~Yuksel,
\PRD {\bf 79}, 055019 (2009).

\bibitem{susy8}
  B.~Kyae,
  \JCAP {\bf 0907}, 028 (2009).

\bibitem{susy9}
  R.~Allahverdi, B.~Dutta, K.~Richardson-McDaniel and Y.~Santoso,
  \PLB {\bf 677}, 172 (2009).

\bibitem{susy10}
  R.~C.~Cotta, J.~S.~Gainer, J.~L.~Hewett and T.~G.~Rizzo,
{\it New J. Phys.} {\bf 11}, 105026 (2009).


\bibitem{susy11}
  I.~Gogoladze, R.~Khalid and Q.~Shafi,
  \PRD {\bf 79}, 115004 (2009).

\bibitem{susy12}
  J.~P.~Hall and S.~F.~King,
 {\it JHEP} {\bf 0908}, 088, (2009).

\bibitem{susy13}
  M.~Berg et al.,
 \JCAP {\bf 0908}, 035 (2009).

\bibitem{susy14}
  G.~Belanger, F.~Boudjema, A.~Pukhov and R.~K.~Singh,
 {\it JHEP} {\bf 0911}, 026 (2009).

\bibitem{susy15}
  C.~Liu,
 \PRD {\bf 80}, 035004 (2009).

\bibitem{susy16}
  D.~Feldman, Z.~Liu, P.~Nath and B.~D.~Nelson,
\PRD {\bf 80}, 075001 (2009).

\bibitem{susy17}
  D.~A.~Demir et al.,
\PRD {\bf 81}, 030519 (2010).

\bibitem{nmssmhooper}
  D.~Hooper and T.~M.~P.~Tait,
 \PRD {\bf 80}, 055028 (2009).

\bibitem{nmssmhooper1}
  W.~Wang, Z.~Xiong, J.~M.~Yang and L.~X.~Yu,
 {\it JHEP} {\bf 0911}, 053  (2009).

\bibitem{nmssmlykken}
  Y.~Bai, M.~Carena and J.~Lykken,
 \PRD {\bf 80}, 055004 (2009).

\bibitem{kk}
  Y.~Bai and Z.~Han,
 \PRD {\bf 79}, 095023 (2009).

\bibitem{kk1}
  D.~Hooper and K.~M.~Zurek,
  \PRD {\bf 79}, 103529 (2009).

\bibitem{kk2}
  C.~R.~Chen et al.,
  {\it JHEP} {\bf 0909}, 078 (2009).

\bibitem{light}
  M.~Pospelov and A.~Ritz,
  \PLB {\bf 671}, 391 (2009).

\bibitem{light1}
  A.~E.~Nelson and C.~Spitzer,
  arXiv:0810.5167 [hep-ph].

\bibitem{light3}
  Y.~Nomura and J.~Thaler,
  \PRD {\bf 79}, 075008 (2009).

\bibitem{light4}
  D.~Feldman, Z.~Liu and P.~Nath,
  \PRD {\bf 79}, 063509 (2009).

\bibitem{light5}
  C.~R.~Chen, F.~Takahashi and T.~T.~Yanagida,
  \PLB {\bf 673}, 255 (2009).

\bibitem{light6}
  T.~Hur, H.~S.~Lee and C.~Luhn,
  {\it JHEP} {\bf 0901}, 081 (2009).

\bibitem{light7}
  M.~Pospelov,
\PRD {\bf 80}, 095002 (2009).

\bibitem{light9}
  L.~Bergstrom et al.,
  \PRD {\bf 79}, 081303 (2009).

\bibitem{light10}
  E.~J.~Chun and J.~C.~Park,
  \JCAP {\bf 0902}, 026 (2009).

\bibitem{light11}
  K.~Hamaguchi, S.~Shirai and T.~T.~Yanagida,
 \PLB {\bf 673}, 247 (2009).

\bibitem{light12}
  K.~J.~Bae et al.,
  \NPB {\bf 817}, 58 (2009).

\bibitem{light13}
  T.~Gehrmann, N.~Greiner and P.~Schwaller,
  arXiv:0812.4240 [hep-ph].

\bibitem{light14}
  L.~Covi and J.~E.~Kim,
 {\it New J. Phys.} {\bf 11}, 105003 (2009).

\bibitem{light15}
  R.~Barbieri, L.~J.~Hall, V.~S.~Rychkov and A.~Strumia,
{\it J. Phys. G} {\bf 36}, 115008 (2009).

\bibitem{light16}
  M.~Ibe, Y.~Nakayama, H.~Murayama and T.~T.~Yanagida,
  {\it JHEP} {\bf 0904}, 087 (2009).

\bibitem{light17}
  C.~Cheung, J.~T.~Ruderman, L.~T.~Wang and I.~Yavin,
\PRD {\bf 80}, 035008 (2009).

\bibitem{light18}
  S.~Cassel, D.~M.~Ghilencea and G.~G.~Ross,
\NPB {\bf 827}, 256 (2010).

\bibitem{light19}
  R.~Essig, P.~Schuster and N.~Toro,
\PRD {\bf 80}, 015003 (2009).

\bibitem{light20}
  K.~Kohri, J.~McDonald and N.~Sahu,
 \PRD {\bf 81}, 023530 (2009).

\bibitem{light22}
  J.~Mardon, Y.~Nomura and J.~Thaler,
\PRD {\bf 80}, 035013 (2009).

\bibitem{light23}
  D.~E.~Morrissey, D.~Poland and K.~M.~Zurek,
 {\it JHEP} {\bf 0907}, 050 (2009).

\bibitem{annihilation}
  M.~Cirelli, A.~Strumia and M.~Tamburini,
 \NPB   {\bf 787}, 152 (2007).

\bibitem{annihilation1}
  E.~Ponton and L.~Randall,
  {\it JHEP} {\bf 0904}, 080 (2009).

\bibitem{annihilation2}
  K.~M.~Zurek,
  \PRD {\bf 79}, 115002 (2009).

\bibitem{annihilation3}
  X.~J.~Bi, P.~H.~Gu, T.~Li and X.~Zhang,
 {\it JHEP} {\bf 0904}, 103 (2009).

\bibitem{annihilation4}
  S.~Khalil, H.~S.~Lee and E.~Ma,
 \PRD {\bf 79}, 041701R (2009).

\bibitem{annihilation5}
  Q.~H.~Cao, E.~Ma and G.~Shaughnessy,
  \PLB {\bf 673}, 152 (2009).

\bibitem{annihilation6}
  E.~Nezri, M.~H.~G.~Tytgat and G.~Vertongen,
  \JCAP {\bf 0904}, 014 (2009).

\bibitem{annihilation7}
   D.~J.~Phalen, A.~Pierce and N.~Weiner,
\PRD {\bf 80}, 063513 (2009).

\bibitem{annihilation8}
  F.~Chen, J.~M.~Cline and A.~R.~Frey,
  \PRD {\bf 79}, 063530 (2009).

\bibitem{annihilation9}
  P.~H.~Frampton and P.~Q.~Hung,
 \PLB {\bf 675}, 411 (2009).

\bibitem{annihilation10}
  M.~Cirelli and A.~Strumia,
 {\it New. J. Phys.} {\bf 11}, 105005 (2009).

\bibitem{annihilation11}
  A.~A.~El-Zant, S.~Khalil and H.~Okada,
  arXiv:0903.5083 [hep-ph].

\bibitem{annihilation12}
  J.~H.~Huh, J.~E.~Kim and B.~Kyae,
 \PRD {\bf 80}, 115012 (2009).

\bibitem{annihilation13}
  I.~Gogoladze, N.~Okada and Q.~Shafi,
\PLB {\bf 679}, 237 (2009).

\bibitem{annihilation14}
  W.~L.~Guo and X.~Zhang,
  \PRD {\bf 79}, 115023 (2009).

\bibitem{annihilation15}
  X.~Calmet and S.~K.~Majee,
\PLB {\bf 679}, 267 (2009).

\bibitem{annihilation16}
   P.~H.~Gu, H.~J.~He, U.~Sarkar and X.~m.~Zhang,
\PRD {\bf 80}, 053004 (2009).


\bibitem{Ibarra08PRL} A. Ibarra and D. Tran, \PRL {\bf 100}, 061301 (2008).

\bibitem{decay}
  C.~R.~Chen, F.~Takahashi and T.~T.~Yanagida,
\PLB {\bf 671}, 71 (2009).

\bibitem{decay1}
  C.~R.~Chen and F.~Takahashi,
  \JCAP {\bf 0902}, 004 (2009).

\bibitem{decay2}
   K.~Hamaguchi, E.~Nakamura, S.~Shirai and T.~T.~Yanagida,
  \PLB {\bf 674}, 299 (2009).


\bibitem{decay3}
  A.~Ibarra and D.~Tran,
  \JCAP {\bf 0902}, 021 (2009).


\bibitem{decay4}
  C.~R.~Chen, M.~M.~Nojiri, F.~Takahashi and T.~T.~Yanagida,
{\it Prog. Theor. Phys.} {\bf 122}, 553 (2009).

\bibitem{decay5}
  E.~Nardi, F.~Sannino and A.~Strumia,
  \JCAP {\bf 0901}, 043 (2009).

\bibitem{decay6}
  K.~Ishiwata, S.~Matsumoto and T.~Moroi,
  \PRD {\bf 79}, 043527 (2009).

\bibitem{decay7}
  M.~Pospelov and M.~Trott,
  {\it JHEP} {\bf 0904}, 044 (2009).


\bibitem{decay8}
 J.~Hisano, M.~Kawasaki, K.~Kohri and K.~Nakayama,
  \PRD {\bf 79}, 043516 (2009).

\bibitem{decay9}
  J.~Liu, P.~F.~Yin and S.~H.~Zhu,
\PRD {\bf 79}, 063522 (2009).

\bibitem{decay10}
  F.~Takahashi and E.~Komatsu,
  arXiv:0901.1915 [astro-ph].

\bibitem{decay11}
 C.~H.~Chen, C.~Q.~Geng and D.~V.~Zhuridov,
 \PLB {\bf 675}, 77 (2009).

\bibitem{decay12}
  K.~Hamaguchi, F.~Takahashi and T.~T.~Yanagida,
  \PLB {\bf 677}, 59 (2009).

\bibitem{decay13}
  X.~Chen,
\JCAP {\bf 0909}, 029 (2009).

 \bibitem{decay14}
  K.~J.~Bae and B.~Kyae,
  {\it JHEP} {\bf 0905}, 102 (2009).

 \bibitem{decay15}
  R.~Essig, N.~Sehgal and L.~E.~Strigari,
  \PRD, 023506 (2009).


\bibitem{decay16}
  S.~Shirai, F.~Takahashi and T.~T.~Yanagida,
 \PLB {\bf 675}, 73 (2009).

\bibitem{decay17}
  K.~Ishiwata, S.~Matsumoto and T.~Moroi,
 {\it JHEP} {\bf 0905}, 110 (2009).

\bibitem{decay18}
   M.~Endo and T.~Shindou,
 {\it JHEP} {\bf 0909}, 037 (2009).


\bibitem{decay19}
  S.~L.~Chen, R.~N.~Mohapatra, S.~Nussinov and Y.~Zhang,
  \PLB {\bf 677}, 311 (2009).


\bibitem{decay20}
  K.~Ishiwata, S.~Matsumoto and T.~Moroi,
  arXiv:0903.3125 [hep-ph].

\bibitem{decay21}
  A.~Ibarra, A.~Ringwald, D.~Tran and C.~Weniger,
 \JCAP {\bf 0908}, 017 (2009).

\bibitem{decay22}
  A.~Arvanitaki et al.,
 \PRD {\bf 80}, 055011 (2009).

\bibitem{decay23}
   S.~Shirai, F.~Takahashi and T.~T.~Yanagida,
\PLB {\bf 680}, 485 (2009).

\bibitem{decay24}
  C.~H.~Chen, C.~Q.~Geng and D.~V.~Zhuridov,
  arXiv:0905.0652 [hep-ph].

\bibitem{decay25}
  N.~Okada and T.~Yamada,
 \PRD {\bf 80}, 075010 (2009).

\bibitem{decay26}
  H.~Fukuoka, J.~Kubo and D.~Suematsu,
 \PLB {\bf 678}, 401 (2009).

\bibitem{decay27}
  C.~H.~Chen,
  arXiv:0905.3425 [hep-ph].

\bibitem{decay28}
  L.~Zhang, G.~Sigl and J.~Redondo,
\JCAP {\bf 0909}, 012 (2009).

\bibitem{decay29}
  C.~H.~Chen, C.~Q.~Geng and D.~V.~Zhuridov,
 \JCAP {\it 0910}, 001 (2009) [arXiv:0906.1646].

\bibitem{decay30}
  D.~Aristizabal Sierra, D.~Restrepo and O.~Zapata,
 \PRD {\bf 80}, 055010 (2009).

\bibitem{decay31}
  J.~H.~Huh and J.~E.~Kim,
 \PRD {\bf 80}, 075012 (2009).

 \bibitem{decay32}
  D.~G.~E.~Walker,
  arXiv:0907.3142 [hep-ph].

 \bibitem{decay33}
  H.~Murayama and J.~Shu,
 \PLB {\bf 686}, 162 (2010).

  \bibitem{decay34}
  W.~Buchmuller et al.,
 \JCAP {\bf 0909}, 021 (2009).

\bibitem{Bahcall80ApJ} J. N. Bahcall and R. M. Soneira, {\it Astrophys. J. Suppl.} {\bf 44}, 73 (1980).

\bibitem{NFW} J. F. Navarro, C. S. Frenk and S. D. M. White, \apj {\bf 462}, 563(1996).

\bibitem{Moore99MN} B. Moore et al., \MNRAS {\bf 310}, 1147 (1999).

\bibitem{Lavalle08AA}
 J.~Lavalle, Q.~Yuan, D.~Maurin and X.~J.~Bi,
  \aaa {\bf 479}, 427 (2008).

\bibitem{Dutta09PRD} B. Dutta, L. Leblond and K. Sinha, \PRD {\bf 80}, 035014 (2009).

\bibitem{sommerfeld}
A. Sommerfeld, {\it Annalen der Physik} {\bf 403}, 257 (1931).

\bibitem{Lattanzi09PRD} M. Lattanzi and J. I. Silk, \PRD {\bf 79}, 083523 (2009).



\bibitem{Nezri09JCAP}  E. Nezri, M. H. G. Tytgat, and G. Vertongen, \JCAP {\bf 0904}, 014 (2009).

\bibitem{Bergstrom09PRL}
    L.~Bergstrom, J.~Edsjo, and G.~Zaharijas,
    \PRL {\bf 103}, 031103 (2009).

\bibitem{Pinzke09PRL} A. Pinzke, C. Pfrommer, and L. Bergstr\"{o}m, \PRL {\bf 103},
181302 (2009).

\bibitem{Cirelli10} M. Cirelli, P. Pancic, P. D. Serpicoa, (arXiv:0912.0663).

\bibitem{Papucci10} M. Papuccia and A. Strumia, (arXiv:0912.4504).

\bibitem{Feng09} J. L. Feng, M. Kaplinghat, H. B. Yu, \PRL {\bf 104}, 151301 (2010).

\bibitem{Bertone09JCAP} G. Bertone, M. Cirelli, A. Strumiac, and
M. Taoso, \JCAP {\bf 0903}, 009 (2009).

\bibitem{Hutsi10} G. H\"utsi, A. Hektor, and M. Raidal, JCAP, submitted, arXiv:1004.2036 (2010)

\bibitem{Lin10} T. Lin, D. Finkbeiner, and G. Dobler, arXiv:1004.0989 (2010)

\bibitem{Ibarra08JCAP} A. Ibarra and D. Tran,
\JCAP {\bf 0807}, 002 (2008).

\bibitem{Covi09JCAP} L. Covi, M. Grefe, A. Ibarra, and D. Tran,
\JCAP {\bf 0901}, 029 (2009).

\bibitem{Chen10JCAP} C. R. Chen, S. K. Mandal, and F. Takahashi,
\JCAP {\bf 1001}, 023 (2010).

\bibitem{Liu10} J. Liu, Q. Yuan, X. J. Bi, H. Li, and X. M. Zhang, (arXiv:0911.1002).

\bibitem{Hooper09PRD-c}
    D. Hooper, A. Stebbins, and K. M. Zurek,
    \PRD {\bf 79}, 103513 (2009).

\bibitem{Ibarra10JCAP} A. Ibarra, D. Tran, and C. Weniger,
\JCAP {\bf 1001}, 009 (2010).

\bibitem{Zhang10} L. Zhang, C. Weniger, L. Maccione, J.
Redondo, and G. Sigl, (arXiv:0912.4504).

\bibitem{CDMS-II} Z. Ahmed, et al., (arXiv:0912.3592).


\end{thebibliography}
\end{document}